\documentclass[12pt]{article}
\usepackage[utf8]{inputenc}
\usepackage[svgnames,dvipsnames]{xcolor}
\usepackage[a4paper,left=1.5cm,right=1.5cm,top=1.5cm,bottom=1.5cm]{geometry}
\usepackage[bottom, multiple]{footmisc}
\usepackage{graphicx}
\usepackage{tikz}
\usetikzlibrary{decorations.markings}
\usetikzlibrary{decorations.pathmorphing}
\usepackage{framed}
\usepackage{csquotes}
\definecolor{shadecolor}{rgb}{0.90,0.90,0.90}
\usepackage{cite}
\usepackage{hyperref}
\usepackage{subcaption}
\usepackage{pifont}
\usepackage{setspace}
\usepackage{amsmath, amssymb, amsthm, float, graphicx,amsfonts,braket}
\numberwithin{equation}{section}
\usepackage{amsthm}

\theoremstyle{definition}

\hypersetup{colorlinks=true, linkcolor=Maroon, citecolor=FireBrick,urlcolor=Green,linktocpage}
\definecolor{amber(sae/ece)}{rgb}{1.0, 0.49, 0.0}
\definecolor{deepsaffron}{rgb}{1.0, 0.6, 0.2}

\usepackage{multirow}

\usepackage{amstext} 
\usepackage{array}   
\newcolumntype{L}{>{$}l<{$}}
\newcolumntype{C}{>{$}c<{$}}

\def\beq{\begin{eqnarray}}\def\eeq{\end{eqnarray}}
\def\be{\begin{equation}}\def\ee{\end{equation}}

\def\p{\pi}

\def\s{\sigma}
\def\m{\mu}

\def\a{\alpha}

\def\k{\kappa}
\def\b{\beta}

\def\th{\theta}
\def\t{\tau}
\def\D{\Delta}
\def\G{\Gamma}
\def\l{\lambda}

\def\ma{{\mathcal{A}}}

\def\mm{{\mathcal{M}}}

\def\G{\Gamma}

\def\mW{{\mathcal{W}}}

\usepackage{bm}
\usepackage{bm}

\begin{document}

\title{\bf Bootstrapping High-Energy Observables}
\date{}
\author{Faizan Bhat$^{a}$\footnote{faizanbhat@iisc.ac.in}, Debapriyo Chowdhury$^{a}$\footnote{debapriyoc@iisc.ac.in}, Aninda Sinha$^{a}$\footnote{asinha@iisc.ac.in} \\ Shaswat Tiwari$^{b}$\footnote{sstiwari@ncsu.edu}~~and Ahmadullah Zahed$^{a,c}$\footnote{azahed@ictp.it}\\
\it ${^a}$Centre for High Energy Physics,
\it Indian Institute of Science,\\ \it C.V. Raman Avenue, Bangalore 560012, India. \\
\it ${^b}$Department of Physics,
\it North Carolina State University,\\ \it 	Raleigh, North Carolina, United States.\\
\it ${^c}$ICTP, International Centre for Theoretical Physics,\\
\it Strada Costiera 11, 34135, Trieste, Italy
}

\maketitle

\abstract{In this paper, we set up the numerical S-matrix bootstrap by using the crossing symmetric dispersion relation (CSDR) to write down Roy equations for the partial waves. As a motivation behind examining the local version of the CSDR, we derive a new crossing symmetric, 3-channels-plus-contact-terms representation of the Virasoro-Shapiro amplitude in string theory that converges everywhere except at the poles. We then focus on gapped theories and give novel analytic and semi-analytic derivations of several bounds on low-energy data. We examine the high-energy behaviour of the experimentally measurable rho-parameter, introduced by Khuri and Kinoshita and defined as the ratio of the real to the imaginary part of the amplitude in the forward limit. Contrary to expectations, we find numerical evidence that there could be multiple changes in the sign of this ratio before it asymptotes at high energies. We compare our approach with other existing numerical methods and find agreement, with improvement in convergence.}

\tableofcontents

\onehalfspacing

\section{Introduction}
The success of the numerical conformal bootstrap program has led to a resurgence of interest in the S-matrix bootstrap--see \cite{SMWPaper} for a recent review. In recent times, the S-matrix bootstrap has been used to examine various processes, including pion scattering \cite{andrea, BHSST, ABAS}, light-by-light scattering \cite{photonbounds}, and even scattering in superstring theory \cite{andreastring}. Much of the focus has been on constraining low-energy data, such as scattering lengths and Wilson coefficients, using the bootstrap.  {In many cases, positivity of the absorptive part of the amplitude leads to bounds on low-energy effective field theories\cite{Ananthanarayan:1994hf, nima, tolley, SCHFlat, RMTZ, ASAZ, PHASAZ, AZ, rattazzi, Wang:2020jxr, yutin2, sasha, Davis:2021oce, vichi, Swamp, FBAZ, Alvarez:2022krz, Fernandez:2022kzi,   Bern:2022yes, Li:2022aby, Hamada:2023cyt, Riembau:2022yse, Albert:2022oes, Du:2021byy, Salcedo:2022aal, Hong:2023zgm, Li:2022rag, Berman:2023jys, Li:2023qzs, Chen:2023bhu, Chen:2023rar, Pozsgay:2023wmy, Hong:2023zgm, Hamada:2023cyt, deRham:2021bll, Acanfora:2023axz, McPeak:2023wmq}--see also \cite{Ma:2023vgc, Albert:2023jtd, Buric:2023ykg, Fardelli:2023fyq, Aoki:2022qbf, Chen:2022nym}}. At high energies, there exists the famous Froissart-Martin bound \cite{froissartmartin}, which says that for theories with no massless particles, the total scattering cross-section in the forward limit cannot grow arbitrarily fast. Probing for such high-energy behaviour for phenomenologically interesting S-matrices using the present formulation of the bootstrap seems out of reach.

In this work, we consider the fully crossing symmetric, massive scalar scattering amplitude, which we denote by $\mathcal{M}(s,t)$. The present formulation of the numerical S-matrix bootstrap \cite{Sboot1, Sboot2, Miro} uses a clever set of variables that map the cut in each channel to the boundary of a disc. Then, using these variables, one writes down a manifestly crossing symmetric basis and imposes partial wave unitarity on the basis. Each basis element goes to a constant in the high-energy limit. Since to do numerics, one is forced to truncate the number of terms in the basis as well as the number of spins, one naturally gets a scattering cross-section that goes to a constant in the high-energy limit. The Froissart-Martin bound indicates that the partial wave spin sum gets effectively truncated, with the maximum spin depending on the centre-of-mass energy. This essentially means that, in order to recover the high-energy behaviour using the bootstrap, one needs to go to very large spins. This is computationally expensive, and while some strategies have been proposed and briefly examined, no clear idea has emerged to allow such studies at high energies.

\begin{figure}[hbt!]
     \centering
     \begin{subfigure}{0.4\textwidth}
         \centering
         \includegraphics[scale=0.45]{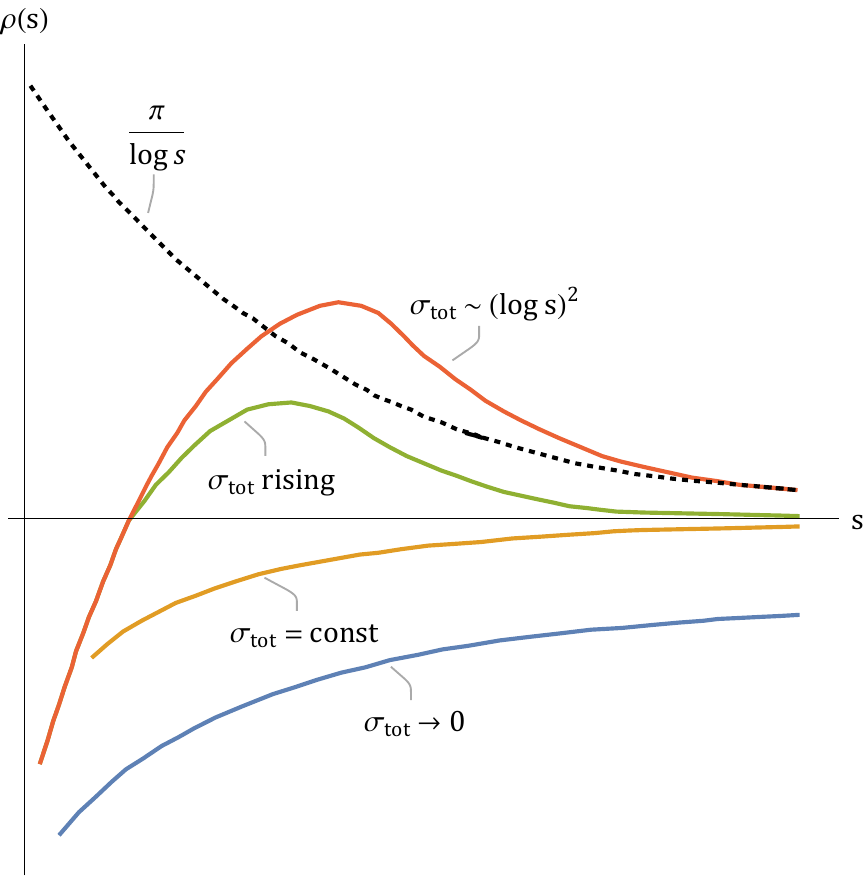}
         \caption{}
\label{rhocartoon}
     \end{subfigure}
     \begin{subfigure}{0.55\textwidth}
         \centering
         \includegraphics[width=0.95\textwidth]{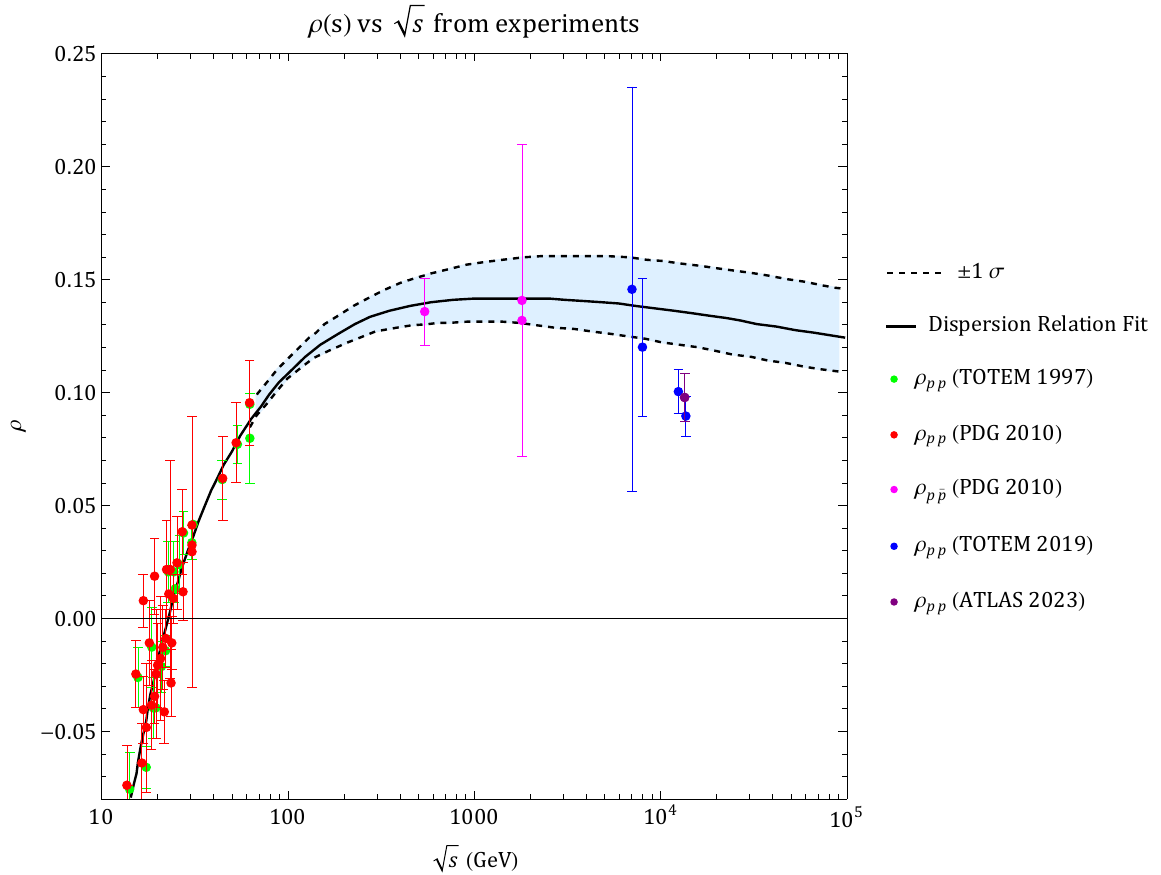}
         \caption{}\label{rhoExpt}
     \end{subfigure}
     \caption{a) Behaviour of $\rho(s)$ \cite{Matthiae:1994uw} for various ranges of $s$.  b) Experimental observations}
\end{figure}

Soon after the Froissart bound was put forth, Khuri and Kinoshita pointed out that the quantity $\rho_{KK}(s)$, defined as the ratio of the real and imaginary parts of the scattering amplitude in the forward limit, can be constrained using ideas in geometric function theory \cite{KK1, KK2}---see \cite{Matthiae:1994uw,Martin:2000bg} for reviews. The $\rho$-parameter can be (and is being) experimentally measured in accelerators\footnote{Recently, \cite{pinn} examined the phase of the S-matrix constrained by elastic unitarity, using Neural Networks.}. We have
\be\label{def:rho}
\rho_{KK}(s)=\frac{\text{Re}[\mathcal{M}(s,0)]}{\text{Im}[\mathcal{M}(s,0)]}\,.
\ee
where we introduce the subscript $KK$ after Khuri-Kinoshita to avoid confusion with other usages\footnote{$\rho$ is used both for the basis variable in the S-matrix bootstrap as well as the $\rho$-meson!} of $\rho$. Khuri and Kinoshita found that for total scattering cross-sections in the forward limit growing like $\sigma_{tot}\sim \log^2 s/s_0$, 
\be\label{expKK}
\rho_{KK}\sim \frac{\pi}{\log s}\,.
\ee
Various possible behaviours \cite{Matthiae:1994uw} of $\rho_{KK}$ depending on how fast $\sigma_{tot}$ grows are depicted in fig.(\ref{rhocartoon}). At the time when this was examined, the experimental measurements in proton-proton scattering had found $\rho_{KK}$ to be negative. The Khuri-Kinoshita prediction, based on the expectation that eventually, the scattering cross-section would rise, was that $\rho_{KK}$ {\it must change sign} at higher energies. This was later borne out, and the latest experimental status is depicted in fig.(\ref{rhoExpt}). In \cite{cerncourier}, Andre Martin described the  $\rho_{KK}$ measurements in the following way: ``This modest experiment in an unglamorous field, deserted by many experimentalists and abandoned by most theoreticians because of calculational difficulties, might be a pointer to new physics, accessible at CERN and at Fermilab.'' It has been pointed out that if $\rho_{KK}$ does not obey one of the expected behaviours, it could indicate the breaking down of one of the assumptions that goes into deriving the dispersive representation (most likely polynomial boundedness) \cite{khurilast}.
The time has come to use modern tools such as the numerical bootstrap to study this quantity in more detail. 

The main goals of this paper are to examine $\rho_{KK}$ using the numerical bootstrap, explore cases where the quantity changes sign, and whether the expected behaviours in fig.(\ref{rhocartoon}) are seen.  We introduce a different way to do numerics (employing the primal optimization approach) using the crossing symmetric dispersion relation (CSDR). While twice subtracted, fixed-$t$ dispersion relations are bread and butter for a theoretical high-energy physicist \cite{SMWPaper, mandelstam, nuss, Martin, ABAS, BHSST, andrea, DualSever, DualKruczenski, DualMiro, piotr, Meltzer, nimayutin2, AnindaKnot, Mizera, Paulos:2020zxx, anant, PHASAZ, Noumi:2022wwf, Haring:2022cyf} (see also \cite{Sinha:2022crx, CarrilloGonzalez:2022fwg}), the CSDR is less familiar. In the standard dispersion relation, one holds $t$-fixed and writes a dispersive integral in $s$. The CSDR, resurrected in \cite{ASAZ}, builds on an old but forgotten paper by Auberson and Khuri \cite{AK}. It maintains manifest crossing symmetry in all three channels at the cost of losing manifest locality. Locality is restored by imposing raints which are equivalent to the so-called ``null constraints" that arise as a consequence of crossing symmetry in the fixed-$t$ scenario. The partial wave expansion resulting from this representation is dubbed the Dyson block expansion \cite{ASAZ} and is spin-wise Regge-bounded. After removing the non-local pieces, one gets an expansion in terms of crossing symmetric partial waves that are local but no longer Regge-bounded. This is reminiscent of Feynman diagrams and includes very important polynomial pieces or contact terms. This is dubbed as the Feynman block expansion \cite{ASAZ}---explicit expressions were first worked out in \cite{DCPHAZ}. In both cases, raints have to be imposed separately. 

Recently, a very compact dispersive representation for the Feynman block expansion was found \cite{Song}, which we refer to as the local CSDR or LCSDR for short. This gives a very compact way of getting the contact terms for any spin. It also enables us to examine the domain of convergence. For instance, by examining the mass-level expansion of the Virasoro-Shapiro amplitude, we find that the Feynman block expansion converges everywhere except at the expected poles, while both the fixed-$t$ and CSDR have finite domains of convergence. 

We use both the CSDR and LCSDR to obtain a different basis than what is being currently used in the literature. This is done by first writing the Roy equation \cite{Roy} using the CSDR/LCSDR which gives us the real part of the partial waves in terms of their imaginary parts. We then only need to make an ansatz for the imaginary part. The next steps are essentially the same: we impose non-linear unitarity on the partial waves using SDPB  \cite{Simmons-Duffin:2015qma} and examine various quantities. We also compare our approach with existing ones in the literature. We introduce some nomenclature, essentially for brevity in writing:
\begin{itemize}
\item The existing approach based on the $\rho$ variable \cite{Sboot1, Sboot2} and extended using wavelet physics ideas \cite{Miro}, will be called $\rho_W$-bootstrap.
\item The Feynman block motivated basis whose compact representation is given via the LCSDR will be called $F_B$-bootstrap.
\item The Dyson block motivated basis will be called $D_B$-bootstrap.
\end{itemize}

For starters, we show that we can reproduce various existing low-energy bounds and also obtain some analytic and semi-analytic results more elegantly. Then we turn our attention to $\rho_{KK}$. In doing so, we find that the $D_B$-bootstrap is better suited for numerics. Rather than maximizing $\rho_{KK}$ itself, which cannot be implemented directly using SDP, we fix some low-energy data such as scattering lengths and extremise the quartic coupling or other scattering lengths. Furthermore, note that the latest LHC data \cite{TOTEM:2019, ATLAS:2022mgx} in fig.(\ref{rhoExpt}) suggests that the $pp$ total cross-section grows slower than the Froissart behaviour. As such, it makes sense to fix low-energy data and let bootstrap determine the high-energy growth.  Fig.(\ref{rhoExpt}) is for $pp$-scattering. It is conventional in the literature to use intuition from pion-pion scattering and apply it here. Since the mass of a proton is approximately $1$GeV, the crossing point is roughly at $s\sim 530$ in these units, while the peak is at $s\sim 10^6$. The Froissart bound indicates that there should be a truncation in spin to $L_{max}\sim \sqrt{s}\log s/s_0$. Assuming $\log s/s_0 \sim O(1)$, and using $L_{max}\sim \sqrt{s}$ as the estimate for the number of partial wave spins, we see that getting the zero reliably will optimistically require approximately $20-30$ spins while getting the peak will require $L \approx 10^3$ spins--- the latter seems beyond the current scope of numerics. In our case, we do the analysis by fixing the spin-0 and spin-2 scattering lengths to the {\it pion experimental values}. In this case, we find that the crossing and the peak lie at much lower values, needing $O(10)$ spins for convergence. In any case, our goal is to study qualitative features of fig. (\ref{rhoExpt}). The kinds of questions that the bootstrap enables us to probe are as follows.
\begin{itemize}
\item Fig.(\ref{rhocartoon}) suggests only one change of sign if any. But can there be multiple sign changes? The high energy values in fig.(\ref{rhoExpt}) do indicate a faster fall-off after the peak than that expected from a dispersion-relation-motivated interpolation (which is trustworthy only at asymptotic energies). 

\item For the cases where there is a change of sign in $\rho_{KK}$, do low-energy observables control the location where this happens? To address this question, we will consider two main scenarios  (i) Fixing spin-0 and spin-2 scattering lengths to the pion values and minimising the coupling (related to $\mathcal{M}(4/3,4/3)$). (ii) Fixing the spin-2 scattering length to the pion value and minimising the spin-0 scattering length. The two results look similar except around $s=4$, which suggests that the spin-2 scattering length is what controls the higher energy behaviour in both cases. We also look at various other interesting possibilities for $\rho_{KK}$ in Appendix \ref{LeafRhos}. In many cases, we find markedly distinct behaviour than what is shown in fig.(\ref{rhocartoon}). 
\item Beyond some large enough $s$, the physics should be sensitive to extra dimensions if they exist. In that case, we should impose higher dimensional unitarity. How does this manifest itself in $\rho_{KK}$? Can the deviation of the experimental values in fig.(\ref{rhoExpt}) be explained using extra dimensions?
\end{itemize}

Fixing the scattering lengths to phenomenological pion values gives faster convergence in spin, needing $O(10)$ spins. There already exist bounds due to Lopez and Mennessier \cite{Lopez:1976zs} from 1976 on the coupling as a function of the spin-2 scattering length, arising from the dual optimization problem. Our primal approach gives good agreement with this remarkable work done almost five decades back, in the absence of Mathematica, SDPB and 40-core workstations \footnote{Needless to say, much of the work in this paper was done using these tools!}.

This paper is organized as follows: In section 2, we study a manifestly crossing symmetric, three-channels-plus-contact-terms representation of the Virasoro-Shapiro amplitude that arises from the LCSDR. This is to motivate the use of CSDR/LCSDR to obtain a basis to set up the numerical bootstrap. In section 3, we review the CSDR and the LCSDR. We use these to obtain some analytic and semi-analytic bounds in section 4. In section 5, we use these to set up numerical S-matrix bootstrap with a new basis. We re-derive many existing results and also provide a fresh derivation of certain old numerical results. In section 6, we turn to $\rho_{KK}$ and examine it using the bootstrap in some detail. We conclude with a discussion of our methods and results, and list some future directions in section 7. The appendices have many helpful intermediate results and details that complement the main text, including a comparison of our techniques with existing numerical methods.

\section{A string theory motivation}

Why should one consider a basis inspired by the crossing symmetric dispersion relation? To motivate this, let's examine the 2-2 scattering amplitude of tachyons in bosonic closed string theory\footnote{This is usually the first amplitude one learns in a string theory course. Our analysis is applicable more generally whenever there is a mass gap. The case of four-dilaton scattering amplitude, which admits a graviton exchange is a bit more subtle and will be considered elsewhere.} which is given by the Virasoro-Shapiro amplitude.
\be
M_{VS}(s,t)=\frac{\Gamma(-1-\frac{s}{4})\Gamma(-1-\frac{t}{4})\Gamma(-1-\frac{u}{4})}{\Gamma(2+\frac{s}{4})\Gamma(2+\frac{t}{4})\Gamma(2+\frac{u}{4})}\,.
\ee
From the worldsheet picture, it comes from
\be
-\frac{1}{2\pi i}\int d^2\sigma |\sigma|^{-\frac{u}{2}-4}|\sigma-1|^{-\frac{s}{2}-4}\,.
\ee
Here $s+t+u=-16$ as the external particles are tachyonic. To go from the worldsheet form to the Gamma-function form, we need to analytically continue. Recently, in \cite{Sen19}, using insights from string field theory, a representation was given whereby the need for analytic continuation was avoided. The result was a representation in terms of a sum over the exchanged mass states, which had contributions from {\it all three} channels, but crucially, needed contact terms. That such a form is possible has been known for a long time and is discussed in chapter 9 of Polchinski's textbook \cite{Polchinski:1998rq}. However, to the best of our knowledge, no explicit form has ever been reported in the literature. The form reported in \cite{Sen19} is not fully explicit since the contact terms there can only be evaluated numerically. Thus, the challenge in front of us is to use dispersion relations to come up with a similar representation that is analytic for any $s,t$ except at the massive poles. We will consider the fixed-$t$ dispersion relation as well as the crossing symmetric dispersion relations one by one, quoting the final answers and comparing with the string field theory form.

Using a fixed-$t$ dispersion relation\footnote{The fixed $t$ dispersive representation we consider (in terms of the shifted variables $s_i$) is
\be
\frac{M\left(s_1, s_2\right)}{\left(b-s_1\right)\left(s_1+s_2+b\right)}=\frac{1}{2 \pi i} \oint_{s_1} \frac{d s_1^{\prime}}{\left(s_1^{\prime}-s_1\right)} \frac{M\left(s_1^{\prime}, s_2\right)}{\left(b-s_1^{\prime}\right)\left(s_1^{\prime}+s_2+b\right)},
\ee
The above contour only contains the pole $s_1$ in the $s_1^{\prime}$-plane. One can open up the contour and pick up the other poles (and branch cuts). We choose $b=0$. }, we get
\be 
\begin{split}
M_{VS}^{(fixed-t)}=& M_{VS}(-\frac{16}{3},t)+\sum_{k=0}^{\infty}\Big[\frac{1}{k-\frac{s}{4}-1}+\frac{1}{k-\frac{u}{4}-1}
-\frac{1}{k+\frac{t}{4} +\frac{5}{3}}-\frac{1}{k+\frac{1}{3}}\Big]\frac{ \left(\left(\frac{t}{4}+2\right)_k\right){}^2}{(k!)^2}\,.
\end{split}
\ee
The notation $(a)_b=\Gamma(a+b)/\Gamma(a)$ is the Pochhammer symbol. This representation is not explicitly crossing symmetric. As such, to reproduce the correct set of $t$-channel poles, it is expected that there is a finite domain of analyticity, which is precisely what happens; as quoted below, the representation does not converge for arbitrary $t$.
For the crossing-symmetric dispersion relation, we find
\be\label{nonlocal}
M_{VS}^{(CSDR)}=\frac{\Gamma(1/3)^3}{\Gamma(2/3)^3}+\sum_{k=0}^{\infty}\left[\frac{1}{k-\frac{s}{4}-1}+\frac{1}{k-\frac{t}{4}-1}+\frac{1}{k-\frac{u}{4}-1}-\frac{3}{k+\frac{1}{3}}\right]\frac{\left(\left(\Lambda ^{(k)}\right){}_k\right){}^2}{(k!)^2}\,,
\ee
where
\be
\Lambda^{(k)}=\frac{1}{6} \left((3 k+1) \left(\sqrt{\frac{12 a}{-3 a+12 k+4}+1}-1\right)+4\right)\,.
\ee
Here $a=y/x$ where $x=-(s_1 s_2+s_1 s_3+s_2 s_3), y=-s_1 s_2 s_3$ with $s_i$'s being the shifted Mandelstam variables $s_1=s+16/3, s_2=t+16/3, s_3=u+16/3$. This representation is manifestly crossing symmetric. While encouraging, some further analysis shows that even this does not converge for all values of $a$--see below. One drawback of this representation, compared to the string field theory form, is that, as discussed in \cite{ASAZ}, this representation is not manifestly local. On expanding around $a=0$, we get negative powers of $x$ for a fixed mass level, which is not expected in a local theory. In \cite{ASAZ, RGASAZ}, it was shown that a manifestly local form is possible after subtracting off the non-local terms. In \cite{DCPHAZ}, hints were found that this local form has a larger domain of convergence. Recently, in \cite{Song}, a local compact dispersive form was worked out, which coincided with the local form proposed in \cite{ASAZ}. The advantage of this local dispersive form is its simplicity--to be reviewed below--which avoids the need to work out the contact terms for each partial wave separately. This form leads to:
\be\label{local}
M_{VS}^{(LCSDR)}(s,t)=\sum_{k=0}^{\infty}\left[\frac{1}{k-\frac{s}{4}-1}+\frac{1}{k-\frac{t}{4}-1}+\frac{1}{k-\frac{u}{4}-1}-\frac{1}{k+\frac{1}{3}}\right]\frac{\left(\left(\Lambda_L ^{(k)}\right){}_k\right){}^2}{(k!)^2}
\ee
where
\be
\Lambda_L^{(k)}=\frac{1}{6} \left((3 k+1) \left(\sqrt{\frac{y}{16(k+\frac{1}{3})^3}+1}-1\right)+4\right)\,.
\ee
Now, remarkably, this representation converges everywhere, much like what is expected from the string field theory form. Despite appearances, at each $k$'th mass level, the combination of Pochhammers simplify to a degree-$k$ polynomial in $y$ with integer zeros at $y<0$ and moreover, the poles in each channel get multiplied by the same, manifestly positive (for real Mandelstam variables) quantity\footnote{The Regge behaviour of this representation is far from obvious but it can be numerically verified that it indeed has the expected form. Another point we note here is that each individual Pochhammer is also positive for $s,t>0$.}. Thus, we now have an explicit analytic answer, unlike \cite{Sen19}. Although not obvious, the residues at the poles coincide in both representations and agree with the Gamma function form. The two representations in eq.(\ref{nonlocal}) and eq.(\ref{local}) also give the same regular pieces or contact terms after removing the non-local terms in eq.(\ref{nonlocal}). The locality/null constraints get mapped to Regge boundedness constraints in the new representation. We summarize the situation with convergence below.
\subsubsection*{Convergence:}
We compute the large $k$ limit of the summand in all three representations and find
\be
\begin{split}
&\text{CSDR} \approx k^{-\frac{11}{3}+\frac{a}{2}}\\
&\text{Fixed-t} \approx k^{-\frac{11}{3}+\frac{s_2}{2}}\\
&\text{L-CSDR} \approx k^{-\frac{11}{3}}\log(k)\,.
\end{split}
\ee
Hence we have convergence for $\Re(a)<\frac{16}{3}$ in the case of CSDR and for $\Re(s_2)<\frac{16}{3}$ in the case of fixed-$t$. The LCSDR converges for any choice of $a$ or $s_2$ as large $k$ is independent of them. Numerical comparison of convergence is given in the table \eqref{tab:VS} in Appendix \ref{StringRedux}. Thus the crossing symmetric dispersion relation, with non-local pieces discarded, promises to have a bigger domain of convergence. Roy and Wanders in 1978 already reported on the possibility of a bigger domain of convergence using a CSDR \cite{roywanders} and in the future, it will be worthwhile to take this line of research to completion. We do not do this here but find the above example a good motivation to set up the numerical S-matrix bootstrap using the (L)CSDR.

\section{CSDR: A brief review}
We consider 2-2 scattering amplitude of identical massive scalars. It is crossing symmetric in all three channels and, when polynomial boundedness and unitarity are obeyed, respects the Froissart-Martin bound \cite{froissartmartin}. These properties, combined with assumptions of the analyticity domain, allow us to derive a dispersive representation that maintains full crossing symmetry \cite{AK, ASAZ, RGASAZ}. Such a representation is not only mathematically elegant but also provides valuable analytical tools for various inquiries within the framework of the bootstrap approach \cite{PHASAZ, PRAS, AZ, Bissi:2022fmj, spinTR, SGPRAS,  tolley2, Alday1, Fardelli:2023fyq, AZ2}. It is given as
\be\label{disper0}
\begin{aligned}
& \mathcal{M}_0\left(s_1, s_2\right)=\alpha_0+\frac{1}{\pi} \int_{\frac{8}{3}}^{\infty} \frac{d \t}{\t} \mathcal{A}_0\left(\t ; \widehat{s}_2\left(\t, \beta \right)\right) \times H_0\left(\t ; s_1, s_2, s_3\right),
\end{aligned}
\ee
where $\a_0 = \mathcal{M}_0(0,0)$ is the subtraction constant, $\mathcal{A}_0\left(s_1 ; s_2\right)$ is the $s$-channel discontinuity and
$$
\begin{aligned}
& H_0(\t ; s_1,s_2,s_3) \equiv \frac{s_1}{\t-s_1}+\frac{s_2}{\t-s_2}+\frac{s_3}{\t-s_3}, \\
& \widehat{s}_2(\t, \beta) \equiv  \t \frac{-1 + \sqrt{1+4 \beta}}{2}, \quad \beta = \frac{a}{\t - a}
\end{aligned}
$$
We will shift between the following two notations as per convenience: $\mathcal{M}(s,t)=\mathcal{M}_0\left(s_1, s_2\right)$, $\mathcal{A}(s,t)=\mathcal{A}_0\left(s_1, s_2\right)$ and $H_0\left(\t ; s_1, s_2, s_3\right)=H\left(s' ; s, t, u\right)$, where we replace $s_1$ by $s -\frac{4}{3}$, $s_2$ by $t- 4/3$ in the arguments and change the integration variable $\t$ to $s' = \t + \frac{4}{3}$. 
Lorentz symmetry allows us to expand the amplitude in a basis of Legendre polynomials $P_\ell\left(z\right)$ in $d = 4$ as follows 
\begin{equation}\label{PWE}
\mathcal{M}(s,t)=32\pi\sqrt{\frac{s}{s-4}}\sum_{\ell=0}^\infty(2\ell+1) f_\ell(s)P_\ell\left(z\equiv 1+\frac{2t}{s-4}\right) \,.
\end{equation}
where $f_{\ell}(s)$ are the partial wave coefficients that can be extracted from the amplitude using orthogonality of the Legendre polynomials. We get 
\begin{equation}
    f_\ell(s) = \frac{1}{32 \pi} \sqrt{\frac{s-4}{s}} \int_{-1}^1 \frac{dz}{2} P_\ell(z) \mathcal{M}\left(s, t \equiv \frac{(s-4)(z-1)}{2}\right)
\end{equation}
\subsubsection*{The Dyson block expansion, Roy equations and locality constraints}
Plugging \eqref{PWE} for the absorptive part in \eqref{disper0}, we get an expansion of the dispersive representation in partial waves. This expansion is referred to as the Dyson block expansion \cite{ASAZ}. Explicitly,
\begin{equation}\label{DB}
\mathcal{M}(s,t)=\alpha_0 + \frac{1}{\pi}\int_{4}^{\infty} \frac{ds'}{s'- \frac{4}{3}} H(s';s,t,u)\left(32\pi\sqrt{\frac{s'}{s'-4}}\sum_{\ell=0}^{\infty}(2\ell+1)\text{Im} f_\ell(s')P_\ell(\sqrt{\xi(s',a)})\right)
\end{equation}
where $\xi(s',a)=\left(\frac{s'-\frac{4}{3}}{s'-4}\right)^2\frac{s'-\frac{4}{3}+3a}{s'-\frac{4}{3}-a}$. We can plug \eqref{DB} into \eqref{PWE} to get the $\text{Re} f_{\ell}(s)$ for any spin $\ell$ in terms of imaginary parts $\text{Re} f_{\ell'}(s)$ for all the spins $\ell'$. This set of dispersive integral equations for the partial waves are known as the Roy equations \cite{Roy}. They will be important for us to set up our numerics as discussed in section 5. \\
The Dyson block expansion, although fully crossing symmetric and spin-wise Regge bounded, comes at the price that it is not local. Consider the following expansion of the amplitude.
\be \label{WCexp}
\mathcal{M}(s,t)=\sum_{p=0,q=0}^{\infty} \mW_{p,q} x^p y^q \,.
\ee
$x$ and $y$ are crossing symmetric polynomials defined previously. This expansion is completely general; all higher-degree crossing invariant polynomials can be expressed in terms of products of $x$ and $y$. The expansion coefficients $\mW_{p,q}$ are called Wilson coefficients. Crucially, in any local QFT, no terms with negative powers of $x$ can appear in the above expansion. One way to see this is that such terms cannot arise from a local Lagrangian consisting of only fields and their derivatives.  Therefore,
\be \label{nullloc}
\text{for any positive integer $p$ and $q > p$}, ~ \mW_{-p,q} = 0. \qquad \text{(Null/locality constraints)}
\ee
We use the terminology ``null'' or ``locality'' constraints interchangeably. In the case of fixed-$t$ dispersion relations (which are local by construction), the same constraints arise on imposing crossing symmetry \cite{tolley, SCHFlat}, as was shown in \cite{ASAZ, RGASAZ}.
The Dyson block expansion leads to the following general formula for the Wilson coefficients \cite{ASAZ}.
\be \label{WCDef}
\begin{aligned}
\mathcal{W}_{n-m, m} & =\int_4^{\infty} \frac{ds'}{s'- \frac{4}{3}} \frac{1}{\left(s'- \frac{4}{3}\right)^{2 n+m}}  32 \pi  \sqrt{\frac{s'}{s'- 4}}\sum_{\ell=0}^{\infty}(2 \ell+1) \text{Im}f_{\ell}\left(s'\right) \mathcal{B}_{n, m}^{(\ell)}(s'), \\
\mathcal{B}_{n, m}^{(\ell)}(s') & =\sum_{j=0}^m \frac{p_{\ell}^{(j)}\left(\xi_0\right)\left(4 \xi_0\right)^j(3 j-m-2 n)(-n)_m}{\pi  j !(m-j) !(-n)_{j+1}}\,,~ p_{\ell}^{(j)}\left(\xi_0\right)=\partial^j C_\ell(\sqrt{\xi})/\partial \xi^j|_{\xi=\xi_0}
\end{aligned}
\ee
where $\xi_0=\dfrac{\left(s'- \frac{4}{3}\right)^2}{(s'-4)^2}$.

\subsubsection*{The local CSDR and the Feynman Block expansion}
Recently, a local crossing symmetric dispersion relation (LCSDR, for short) was derived in \cite{Song}. The representation is derived by subtracting the non-local terms from the CSDR. It is given by
\be\label{eq:LCSDR}
\mathcal{M}_0(s_1,s_2)=\alpha_0+\frac{1}{\pi} \int_{\frac{2\mu}{3}}^{\infty} \frac{d \t}{\t} \left(\mathcal{A}_0\left(\t, \widehat{s}_2\left(\t,\eta\right)\right)H_0\left(\t ; s_1, s_2, s_3\right)+ 2\left( \mathcal{A}_0\left(\t, \widehat{s}_2 \left(\t,\eta\right)\right)-\mathcal{A}_0(\t, 0)\right)\right)
\ee 
where $\eta=-\dfrac{s_1 s_2 s_3}{\t^3}$. The partial wave expansion of the LCSDR consists of the three-channel exchange terms and an additional infinite sum of contact terms which are polynomials in $s,t,u$. The expansion was called the Feynman block expansion in \cite{ASAZ} and is given by
\begin{equation}
\mathcal{M}(s,t)=\alpha_0+\frac{1}{\pi}\sum_{\ell=0}^\infty\int_{4}^{\infty}\frac{ds'}{\left(s'-4\right)^\ell}32\pi\sqrt{\frac{s'}{s'-4}}(2\ell+1)\text{Im} f_\ell(s')M_\ell^F(s';s,t)
\end{equation}
where the Feynman blocks are
\begin{equation}
M_\ell^F(s';s,t)=\sum_{i=1}^3M_\ell^{(i)}(s';s,t)+M_\ell^{(c)}(s';s,t)
\end{equation}  
with
\begin{equation}
\begin{split}
&M_\ell^{(1)}(s';s,t)=\left(s-4\right)^\ell P_\ell\left(1+\frac{2t}{s-4}\right)\left(\frac{1}{s'-s}-\frac{1}{s'-\frac{4}{3}}\right)\,,\\
&M_\ell^{(2)}(s';s,t)=M_\ell^{(1)}(s';t,u)\,,M_\ell^{(3)}(s';s,t)=M_\ell^{(1)}(s';u,s)
\end{split}
\end{equation}
and $M_\ell^{(c)}(s';s,t)$ are the contact terms. Their explicit forms were derived first in \cite{DCPHAZ} by subtracting non-local terms block-by-block from the Dyson blocks. However, the LCSDR offers a compact and simpler way to get them directly.



\section{Some analytic/semi-analytic bounds}
In this section, we list some simple results that can be derived using the CSDR/LCSDR. We present some analytic bounds that follow from the positivity properties of Legendre polynomials and linear unitarity ($\text{Im}f_{\ell}(s)>0$), as well as some semi-analytic bounds which require input from bounds obtained from the numerical bootstrap. Let us begin with a theorem that is known using the fixed-$t$ dispersion relation \cite{Martin:1970jsp}.
\\
{\textbf{Theorem:}} \emph{If the scattering amplitude is real, then the subtracted scattering amplitude $\mathcal{M}(s,t) - \mathcal{M}\left(\frac{4}{3}, \frac{4}{3}\right)$ is non-negative, i.e.  $\mathcal{M}(s,t) - \mathcal{M}\left(\frac{4}{3}, \frac{4}{3}\right) \geq 0$ for $s<4$, $t<4$, and $u<4$.}
\\
{\textbf{Proof:}} The scattering amplitude is known to be real in the domain defined by $s<4$, $t<4$, and $u<4$. This domain forms a triangular region in the $t$-$s$ plane, which we will call the \textit{reality triangle}.
In the Dyson block expansion, as expressed in equation \eqref{DB}, the Legendre polynomials that appear are always positive when $-\frac{8}{9}<a<\frac{8}{3}$ as proven in \cite{PHASAZ}. This positivity property can be attributed to the fact that $\sqrt{\xi(s',a)}\geq 1$ for $-\frac{8}{9}<a<\frac{8}{3}$ and $s'> 4$.\\
To prove the theorem, we only need to determine the common domain in which the kernel $H(s',s,t)$ is positive and $-\frac{8}{9}<a<\frac{8}{3}$. Interestingly, this is always true within the reality triangle for $s'>4$. Positivity of the absorptive partial wave coefficients then ensures that the full integral in equation \eqref{DB} is always non-negative. The theorem can also be obtained using the LCSDR. 

{\subsubsection*{Semi-analytic upper bound on $\mW_{1,0}$}}
\begin{figure}[H]
\centering
\includegraphics[scale=0.8]{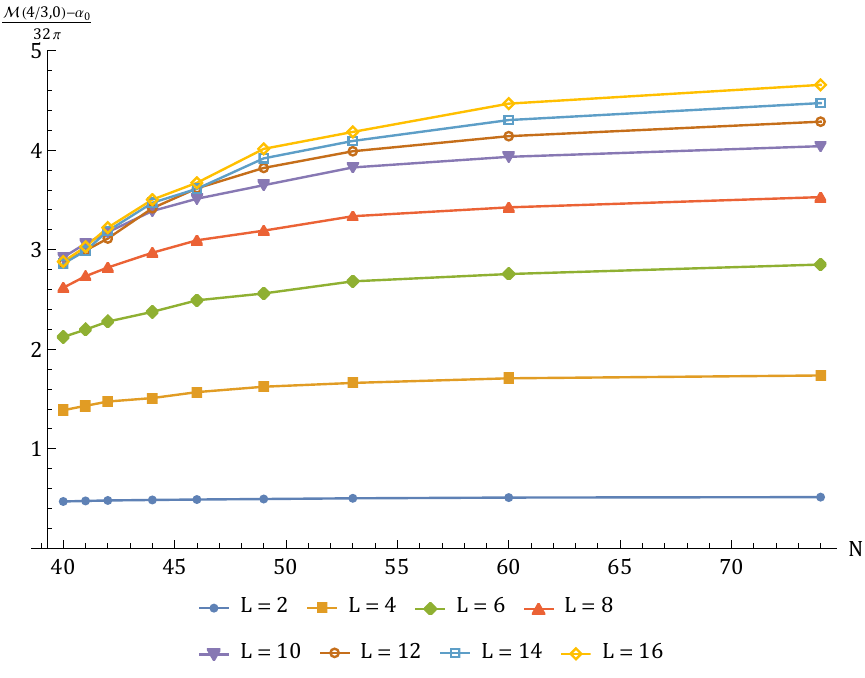}
\caption{Bound from the $D_B$-bootstrap. Unitarity was imposed up to $L =16$ and $s_{max} \approx 4950$. Null constraint $-10^{-12}\leq \mW_{-1,2}\leq 0$ was imposed.}
\label{Combination}
\end{figure}
The bound on $\mW_{1,0}$ can be derived from the fact that the following inequality always holds
\be
\mathcal{M}\left(\frac{4}{3},0\right)-\mathcal{M}\left(\frac{4}{3},\frac{4}{3}\right)=\frac{32}{\pi}\int_{4}^{\infty} ds  \frac{\mathcal{A}\left(s,\frac{4}{3}\right)}{9(s-\frac{4}{3})^3-16(s-\frac{4}{3})}>\frac{32}{\pi} \int_{4}^{\infty} ds \frac{ \mathcal{A}\left(s,\frac{4}{3}\right)}{9\left(s-\frac{4}{3}\right)^3}=\frac{16 \mathcal{W}_{1,0}}{9}\,.
\ee
Now, we can use the bounds $\frac{\mathcal{M}\left(\frac{4}{3},0\right)}{32 \pi} < 3.16$\cite{Lopez:1976zs}  and $\frac{\mathcal{M}\left(\frac{4}{3},\frac{4}{3}\right)}{32\pi} > - 8.08$ \cite{DualSever}  which gives 
$
\frac{\mathcal{W}_{1,0}}{64\pi} < 3.161
$. We also bound the LHS directly using the numerical bootstrap. Figure \eqref{Combination} shows our results (using the $D_B$-bootstrap to be elaborated on in the next section; the $\rho_W$-bootstrap gives a similar result). Imposing unitarity up to spin 16 and one null constraint, we find $\frac{\mathcal{M}\left(\frac{4}{3},0\right)-\mathcal{M}\left(\frac{4}{3},\frac{4}{3}\right)}{32\pi}<4.65$. As can be seen in figure \eqref{Combination}, this bound is close to converging but has not fully converged. So the actual bound must be bigger than this. Nevertheless, this leads to \be\frac{\mathcal{W}_{1,0}}{64\pi} < 1.31,\ee close to the bound $0.93$ found in \cite{Miro} by directly maximising $\mW_{1,0}/(64 \pi)$.

{\subsubsection*{Semi-analytic upper bound on $\mW_{n,0}$}}
\begin{figure}[H]
\centering
\includegraphics[scale=0.8]{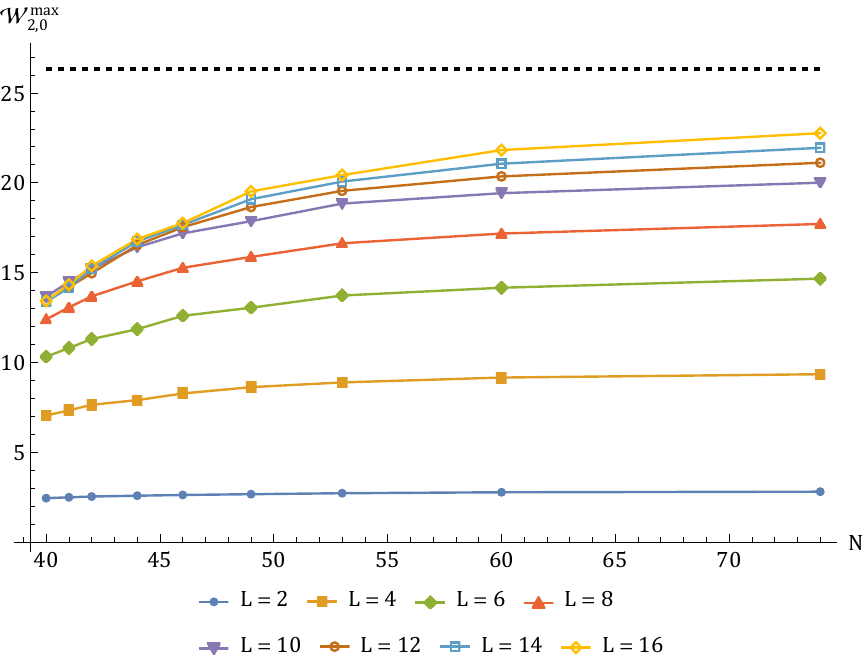}
\caption{Upper bound on $\mathcal{W}_{2,0}$ from the $D_B$-bootstrap. Unitarity was imposed up to $L =16$ and $s_{max} \approx 4950$. Null constraint $-10^{-12}\leq \mW_{-1,2}\leq 0$ was imposed. Black dashed line corresponds to the bound from \eqref{Wn0Bound}.}
\label{W20maxDyson1null}
\end{figure}

The expression for the Wilson coefficients $\mW_{n,0}$ can be written in a simplified form as
\be
\mW_{n, 0} = \frac{2}{\pi}\int_{\frac{8}{3}}^\infty \frac{d\tau}{\tau} \text{Im}f_{\ell}\left(\t+ \frac{4}{3}\right) P_{\ell}\left(\frac{\tau}{\tau-\frac{2\mu}{3}}\right)\left(\frac{1}{\tau^2}\right)^n\,.
\ee
Now using the fact that throughout the integration range, the Legendre polynomial is always positive and $\frac{1}{\t^2}  < \left(\frac{3}{8} \right)^{2}$, and $\frac{\mW_{1,0}}{64\pi} < 0.93$\cite{Miro}, we can write
\be
\label{Wn0Bound}
0 \leq \, \frac{\mathcal{W}_{n, 0}}{64\pi}\, \lesssim \,  0.93\times \left(\frac{3}{8}\right)^{2n-2}\,.
\ee
This gives
$
\frac{\mathcal{W}_{2, 0}}{64\pi}\, \lesssim \,  0.131\,, ~\, \frac{\mathcal{W}_{3, 0}}{64\pi}\, \lesssim \,  0.0184\,,~\, \frac{\mathcal{W}_{10, 0}}{64\pi}\, \lesssim \,  2.00 \times 10^{-8}\,,~ \frac{\mathcal{W}_{100, 0}}{64\pi}\, \lesssim \,  4.23 \times 10^{-85}\,.
$
\\
Using the numerical bootstrap, we check these bounds for the case of $\mathcal{W}_{2, 0}$ as depicted in figure \eqref{W20maxDyson1null}. We find 
\be \frac{\mathcal{W}_{2, 0}}{64 \pi} < \frac{22.76}{64 \pi} = 0.113 \ee
Again, since the bound hasn't fully converged, the actual bound must be slightly bigger. This bound is in close agreement with the bounds arising from positivity/typically-real considerations \cite{tolley, SCHFlat, PHASAZ, PRAS, spinTR} on $\mW_{2,0}/\mW_{1,0}$.


\subsubsection*{Rigorous lower bound on $\mW_{n-1,1}$}
\begin{figure}[hbt!]
     \centering
     \begin{subfigure}{0.47\textwidth}
         \centering
         \includegraphics[width=.9\textwidth]{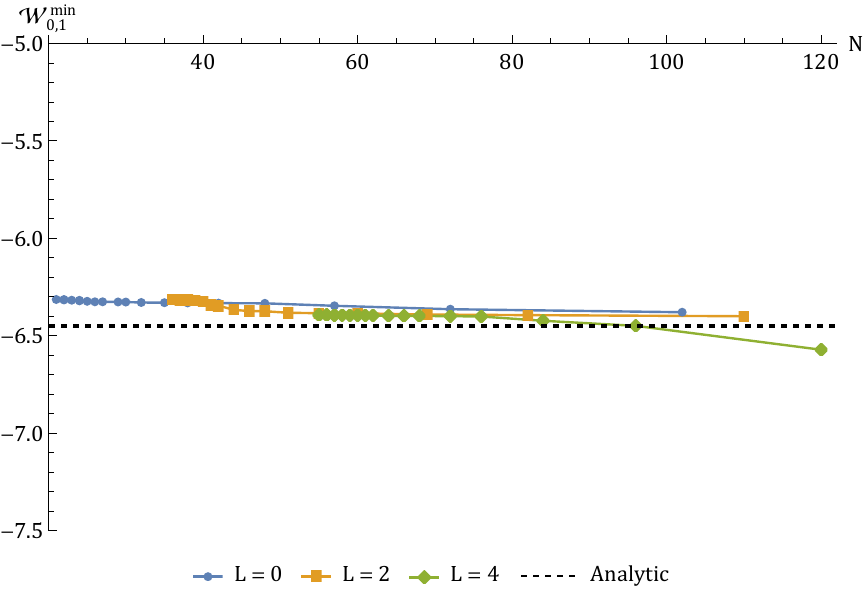}
         \caption{}
     \end{subfigure}
     \begin{subfigure}{0.47\textwidth}
         \centering
         \includegraphics[width=0.9\textwidth]{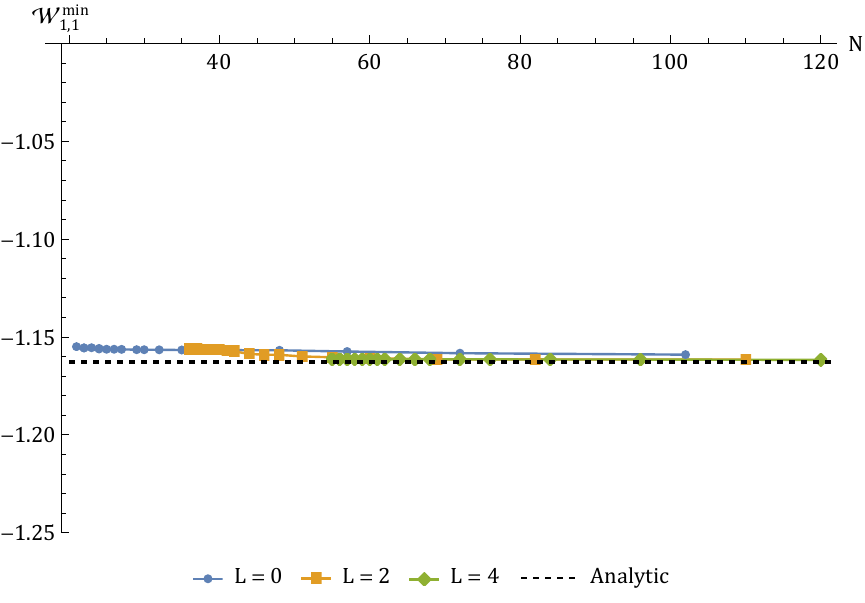}
         \caption{}
     \end{subfigure}
     \caption{Lower bounds on $\mW_{0,1}$ and $\mW_{1,1}$ from the $F_B$-bootstrap. Unitarity was imposed up to $L =4$ and $s_{max} \approx 265$. Black dashed lines correspond to the bound from \eqref{Wnm11Bound}.}
    \label{fig:Wn0LB}
\end{figure}

From the inversion formula for $W_{n-1,1}$, we can show that the spin 0 contribution is always negative while all higher-spin contributions are positive. Thus, keeping only the spin 0 contribution, we get the inequality
\be 
\mW_{n-1,1} \geq -\int_{\frac{8}{3}}^{\infty} \frac{d \t}{\t^{2n+2}} \left(32 (2 n+1) \sqrt{\frac{12}{3 \t-8}+1}  \right )\,,
\ee
where we have put $\text{Im} f_{\ell=0}(s)=1$. After performing the integral, we get in terms of the regularized hypergeometric function ${}_2 \tilde F_1$,
\be 
\label{Wnm11Bound}
\mW_{n-1,1}\geq -12\sqrt{\pi} \left(\frac{3}{8}\right)^{2n}\, \Gamma(2n+2)_2\widetilde F_1\left(-\frac{1}{2},2 n+1;2 n+\frac{3}{2};-\frac{1}{2}\right)\,.
\ee
For example,
$
\mW_{0,1}\gtrsim-6.4514\,,~\mW_{1,1}\gtrsim-1.1627\,,~ \mW_{10,1}\gtrsim -5.32 \times 10^{-8}\,,~ \mW_{100,1}\gtrsim -3.34 \times 10^{-84}
$. When $n\gg 1$, the RHS asymptotes to $-24\sqrt{2\pi n} (3/8)^{2n+1/2}$, which is an exponential suppression. In figure \eqref{fig:Wn0LB}, we show the results from the numerical bootstrap. The plots seem to suggest that the numerical bounds converge to the numbers above, suggesting that these bounds are optimal.

\section{Numerics: A new take on old results}
The S-matrix bootstrap has seen exciting advancements thanks to the numerical implementation of unitarity, crossing symmetry, and analyticity. The problem of optimizing any quantity given some constraints can be viewed from two different approaches - \textit{primal} and \textit{dual}. The primal approach involves finding bounds by scanning the space of amplitudes that are \textit{allowed} by the bootstrap constraints. Since, in practice, we can only impose a finite set of constraints, it means that the bounds are not rigorous, as adding new constraints may disallow some of the amplitudes that were allowed before. On the other hand, the dual approach involves finding bounds by scanning the space of amplitudes that are \textit{disallowed} by the bootstrap constraints. This leads to rigorous bounds as any amplitude, once disallowed, cannot be ruled in by adding new constraints. Since the dual and the primal formulation approach the boundary of the allowed space of amplitudes from two different sides, the boundary must rigorously lie somewhere in the gap between the two, called the \textit{duality gap}. Most recent studies have focused on the primal approach. The dual problem in $d =2 $ was solved in \cite{Dual2D} while $d>2$ has proven to be more challenging to set up (see \cite{DualKruczenski, DualSever, DualMiro} for some recent works). 

\begin{figure}[hbt!]
     \centering
     \begin{subfigure}{0.47\textwidth}
         \centering
         \includegraphics[width=.9\textwidth]{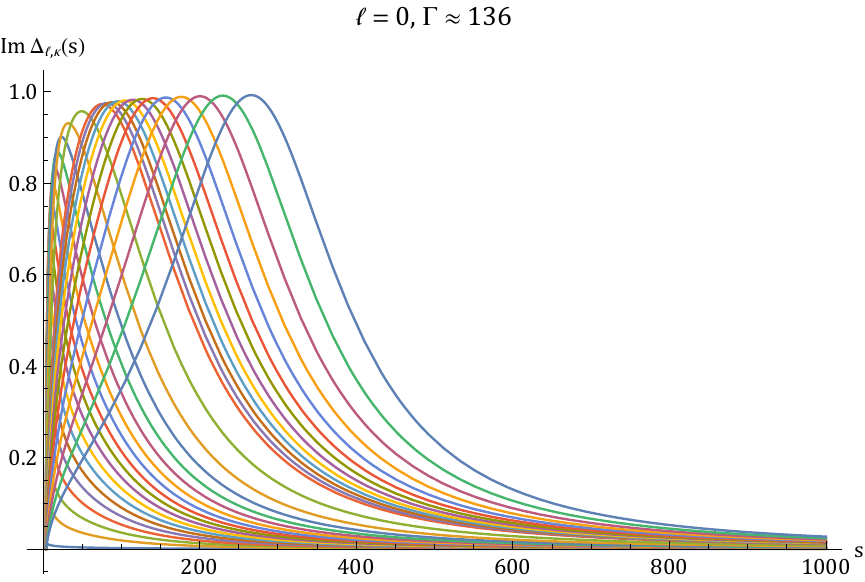}
         \caption{}
     \end{subfigure}
     \begin{subfigure}{0.47\textwidth}
         \centering
         \includegraphics[width=0.9\textwidth]{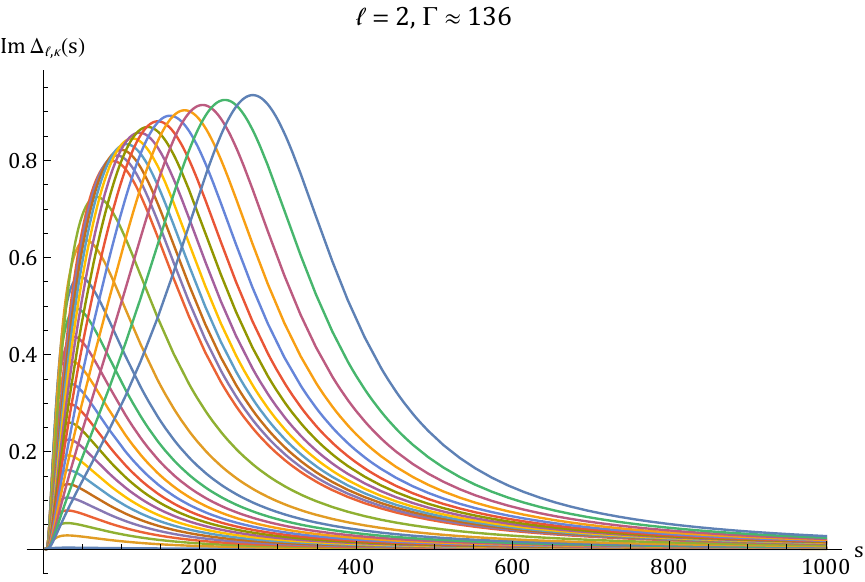}
         \caption{}
     \end{subfigure}
     \caption{Ansatz for the imaginary part of spin 0 and spin 2 partial waves.}
     \label{fig:Wavelets}
\end{figure}

In this paper, we set up the numerical S-matrix bootstrap in the primal approach using the CSDR/LCSDR. Contrary to the previous approaches where a crossing symmetric ansatz is made directly for the amplitude, in our approach, we use the CSDR/LCSDR to write down Roy equations that express the real part of the partial waves in terms of their imaginary part (as discussed in section 3). This naturally leads to a crossing symmetric basis for the amplitude. Our basis assumes maximal analyticity, i.e. it is analytic everywhere apart from a cut along the positive real axis starting at $s=4$ and its images under crossing symmetry. We do not include any bound state poles, but our approach allows for them. Further, in contrast to the previous approaches, a simplifying feature of our approach is that we only need to specify an ansatz for $\text{Im} f_\ell(s)$, which are one-variable functions. In section 2, we examined the crossing symmetric 3-channel representation of the tree level Virasoro-Shapiro string amplitude. This representation is the (infinitely) narrow-width approximation, where the lifetime of the massive states is infinite. We expect loops to broaden the widths. This motivates representing the absorptive part of the partial waves using a Breit-Wigner type basis.  
Based on this line of reasoning, we assume the following ansatz
\begin{equation}
\label{ansatz}
\text{Im} f_\ell(s)=b_{0} \delta_{\ell,0} + \sum_{\kappa \in \Sigma}b_{\ell,\kappa}\:\text{Im}\,\D_{\ell,\kappa}(s), \quad s \geq 4
\end{equation}
where $\D_{\ell,\kappa}(s)= \left(\frac{s-4}{s}\right)^{2\ell+\frac{1}{2}} \frac{\G}{(s-\k)^2+\G}\,\rho_{\k}(s)$, $  \rho_{\k}(s) = \frac{\sqrt{4-s}-\sqrt{\k-4}}{\sqrt{4-s}+\sqrt{\k-4}}$ and 
\begin{equation}
\label{Deltadef}
\begin{split}
\text{Im}\,\D_{\ell,\k}(s)= \left(\frac{s-4}{s}\right)^{2\ell+\frac{1}{2}}\sin \left(2 \arctan \sqrt{\frac{s-4}{\k -4}} \right) \frac{\G}{(s-\k)^2+\G}\,.
\end{split}
\end{equation}
The ansatz is chosen so that the amplitude has an $s$-channel discontinuity starting at $s =4$. We add an additional parameter $b_0$ to the spin 0 absorptive partial wave. Having this factor allows for an infinite spin 0 scattering length (see section $5.2$ for more). Further, the threshold behaviour is fixed by demanding that elastic unitarity is satisfied at $s = 4$ for all partial waves \cite{toolkit}. This is what leads to the explicit $\left(\frac{s-4}{s}\right)^{2\ell+\frac{1}{2}}$ factor. We have an additional ``Breit-Wigner wavelet" type factor $\frac{\G}{(s-\k)^2+\G}$. This introduces peaks in the ansatz when $s = \k$ and $\Gamma$ controls the width of the peak. The motivation to introduce this is to capture the low-lying resonance peaks in the partial waves. Figure \eqref{fig:Wavelets} illustrates the form of the wavelets we use for spin 0 and spin 2 partial waves.
\\
The partial wave coefficients must satisfy unitarity, which in our convention reads $|f_{\ell}(s)|^2 \le \text{Im} f_{\ell}(s) \le 1$. In practice, we impose it via the following positive semi-definiteness condition:
\be
\label{eq:uni}
\begin{pmatrix}
1 + 2\text{Re}f_\ell(s) & 1 - 2\text{Im}f_\ell(s) \\
1 - 2\text{Im}f_\ell(s) & 1 - 2\text{Re}f_\ell(s)
\end{pmatrix} \succeq 0, \quad s \ge 4\,.
\ee
Additionally, as mentioned in Section 2, while the CSDR respects crossing symmetry, it is not local. We must therefore impose locality constraints. These are imposed on the partial wave coefficients using the general formula \eqref{WCDef}. 

We impose the unitarity and locality constraints numerically using the package SDPB\footnote{Recent versions of Mathematica can do SDP up to machine precision. We examined this and concluded that machine precision is not sufficient for our purposes, unfortunately.}. We only impose these constraints up to some maximum spin $L$ and along a grid of values for $s$ up to some $s_{max}$. We also discretize the variable $\k$, which controls where the wavelets (or, more precisely, their peaks) are placed. In general, we find that to get convergent bounds when we impose unitarity up to some large $s_{max}$, we also need to place the wavelets up to some large enough value of $\k$. This is intuitive because the wavelets placed at large $\k$ are expected to capture the properties of the partial waves at large $s$. We provide more details about the numerics in Appendix \ref{NumDetails}. We also compare our bounds and numerics with that of \cite{Miro} in Appendix \ref{Comparison}.

\subsubsection*{Numerical bootstrap using the CSDR or the LCSDR?}
\begin{figure}[hbt!]
     \centering
     \begin{subfigure}{0.49\textwidth}
         \centering
         \includegraphics[width=0.9\textwidth]{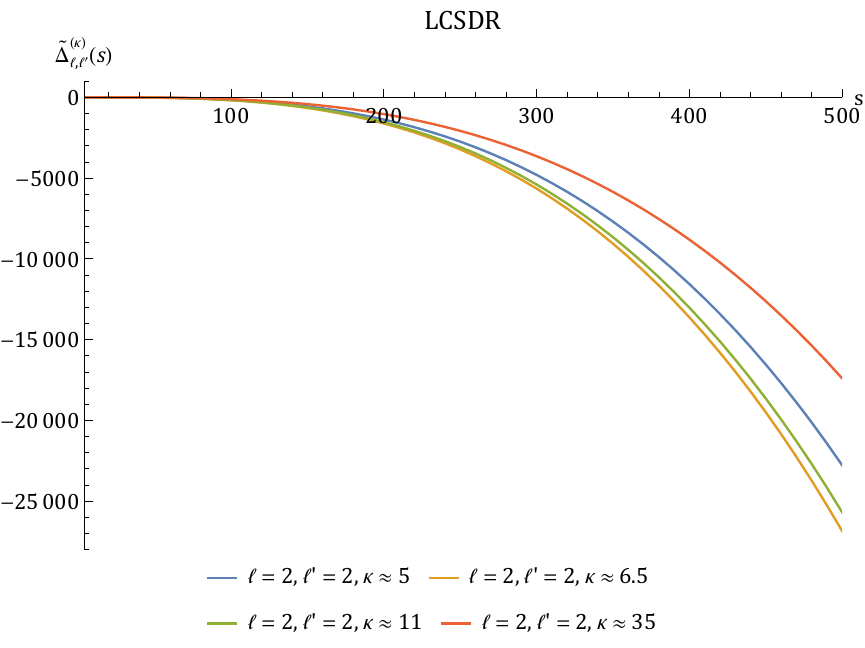}
         \caption{}
     \end{subfigure}
     \begin{subfigure}{0.49\textwidth}
         \centering
         \includegraphics[width=0.9\textwidth]{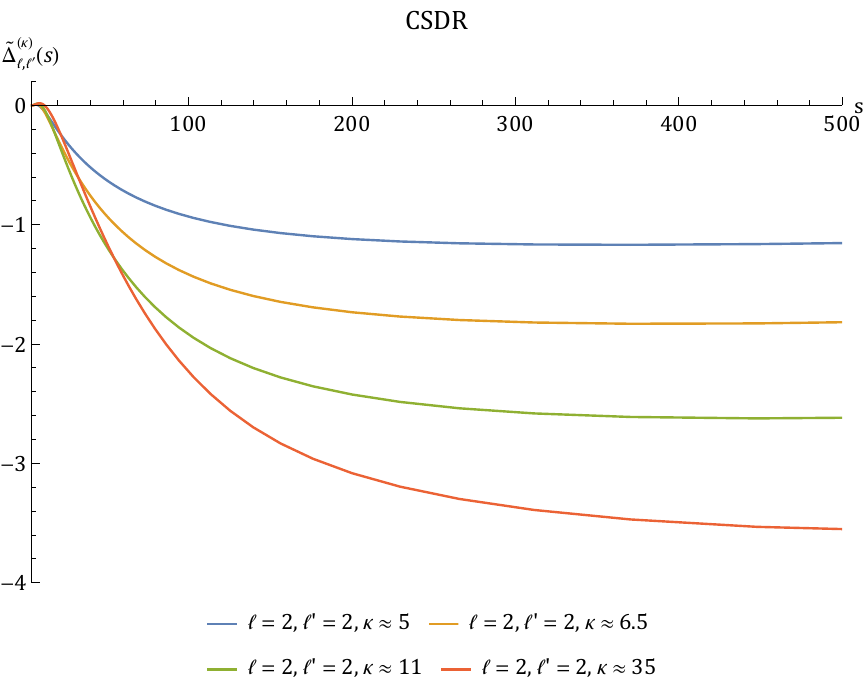}
         \caption{}
     \end{subfigure}
     \caption{Ansatz for the real part of spin two partial waves using the LCSDR and the CSDR.}
     \label{fig: Regge growth}
\end{figure}

We can set up the numerics using both the CSDR and the LCSDR, which we call the $D_B$-bootstrap and $F_B$-bootstrap respectively. A feature of the local representation is that if we truncate the spin sum up to some $\ell = L$, the amplitude is not Regge-bounded but grows polynomially with $s$ due to the contact terms. A sum over infinite spins is therefore required to get the correct Regge behaviour. Since, while doing numerics, we always have to truncate up to some spin, using the LCSDR, it is hard (requires a lot of parameters) to impose unitarity for higher spins for a fixed value of $s_{max}$. It may be possible to ameliorate this by imposing some extra Regge-boundedness conditions but we could not think of an efficient way to implement this and leave it for future investigation. The Dyson block/non-local representation however, always has a bounded Regge growth, making it more suitable to impose unitarity for higher spins. In practice, therefore, we can use the LCSDR to place bounds only on observables that exhibit convergence with fewer spins (low-spin dominance). The CSDR/Dyson block representation is applicable everywhere. Figure \eqref{fig: Regge growth} illustrates the growth of the basis functions $\tilde{\Delta}^{(\kappa)}_{\ell, \ell'} (s)$ that multiply the parameters $b_{\ell, \kappa}$ in the expression for $\text{Re} f_{\ell}(s)$ (after performing the dispersive $s'$ and the $z$ integral) while using the CSDR and the LCSDR. 

In the next section, we present the numerical bounds on various observables using our approach.
\subsection{Bounds on the quartic coupling}
The quartic coupling $\lambda$ is defined as the value of the scattering amplitude at the crossing symmetric but unphysical point.  $\lambda = \frac{1}{32 \pi}\mathcal{M}\left(\frac{4}{3},\frac{4}{3}\right)=\frac{\alpha_0}{32\pi}$, in the notation of eq.(\ref{DB}). 
\begin{figure}[hbt!]
\centering
\includegraphics[scale=0.8]{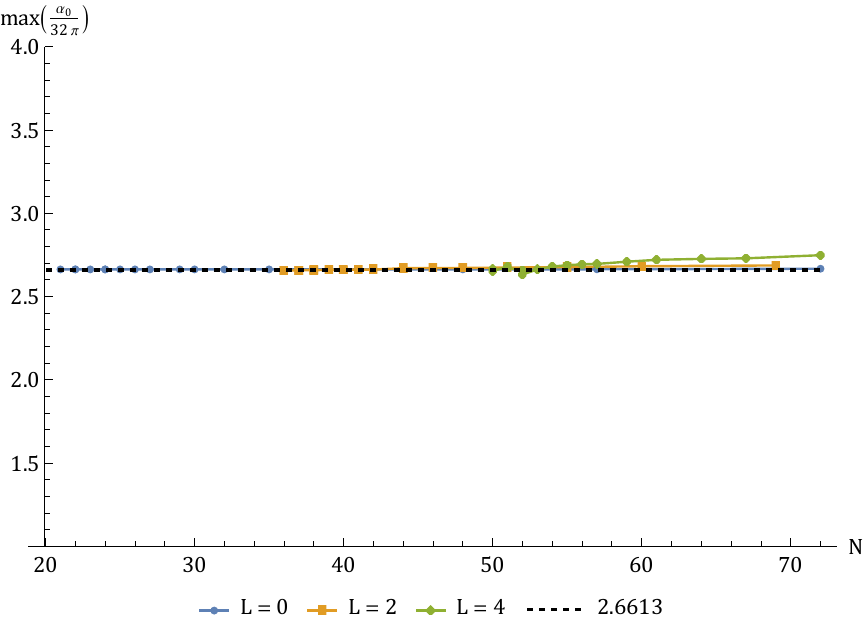}
\caption{Upper bound on $\lambda$ from the $F_B$-bootstrap. Unitarity was imposed up to $L = 4$ and $s_{max} \approx 265$. Black dashed line corresponds to the dual bound from \cite{DualKruczenski}.}
\label{max_con}
\end{figure}
\\
In figure \eqref{max_con}, we illustrate the convergence of the upper bound for the quartic coupling using the LCSDR. From the plot, one can see that imposing unitarity on the spin 0 partial wave alone is sufficient to get the bound. In contrast, the lower bound takes a large number of spins to converge and is not feasible using the $F_B$-bootstrap. Figure \eqref{min_con} shows convergence using the $D_B$-bootstrap. 
\begin{figure}[hbt!]
\centering
\includegraphics[scale=0.8]{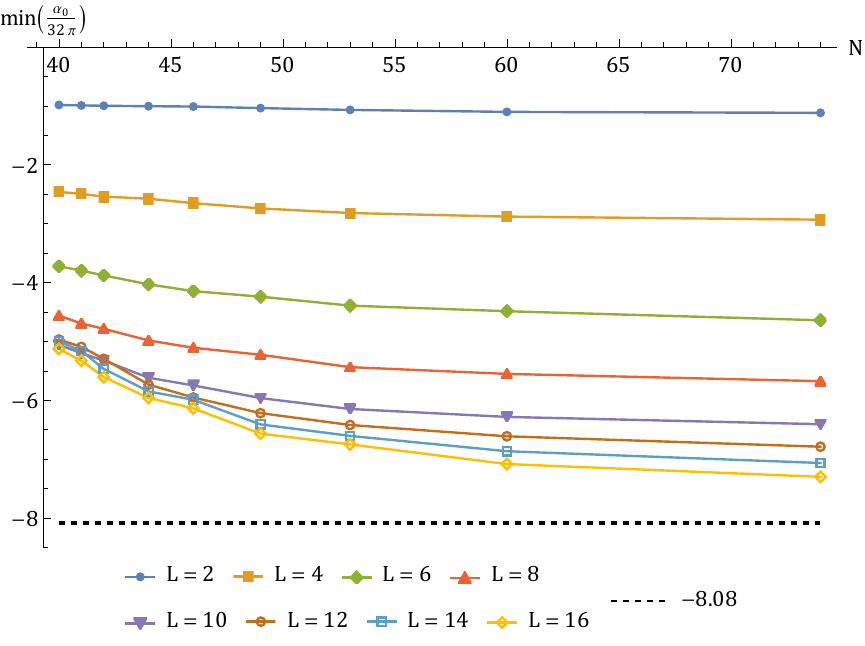}
\caption{Lower bound on $\lambda$ from the $D_B$-bootstrap. Unitarity was imposed up to $L =16$ and $s_{max} \approx 4950$. Null constraint $-10^{-12}\leq \mW_{-1,2}\leq 0$ was imposed.  Black dashed line corresponds to the dual bound from \cite{DualSever}.}
\label{min_con}
\end{figure}

It is more interesting to study bounds on the quartic coupling as a function of the scattering lengths. These are experimentally measurable quantities that capture the behaviour of the partial waves near the threshold. They are defined via\footnote{We have converted to our conventions and given directly the formula in terms of $f_\ell(s)$.} \cite{andrea}:
\be
\text{Re}f_\ell(s) = \left(\frac{s-4}{4}\right)^{\ell+\frac{1}{2}}\left(a_\ell+ \left(\frac{b_\ell}{4}-\frac{a_\ell}{8}\right)(s-4)+\cdots\right)\,,
\ee
where $a_{\ell}$ are the spin-$\ell$ scattering lengths and $b_\ell$ are the corresponding effective ranges. $a_0$ is called the S-wave scattering length, $a_2$ the D-wave scattering length and so on.  In Appendix \ref{ScatteringLengths}, eq. (\ref{inta0eq}), we derive the inequality $a_0\geq \alpha_0/(32 \pi)$.

In figure \eqref{alpha0vsa2}, on the left, we present bounds on the quartic coupling $\lambda$ with $a_2$ fixed to the value for pion scattering, namely $a_2 = a_2^{(pion)} = 0.00175$ \cite{colangelo}. The plot on the right gives a zoomed-in version for small $a_2$ for comparison with \cite{Lopez:1976zs}. In practice, we use the following dispersive representation of $a_2$ given by Yndurain \cite{yndurain}
\be\label{lam2scatt}
a_2=\frac{1}{30\pi^2}\int_{4}^{\infty} \frac{\mathcal{A}(s,4)}{s^3} ds\,.
\ee
\begin{figure}[hbt!]
     \label{alpha0vsa2}
     \centering
     \begin{subfigure}{0.47\textwidth}
         \centering
         \includegraphics[width=1\textwidth]{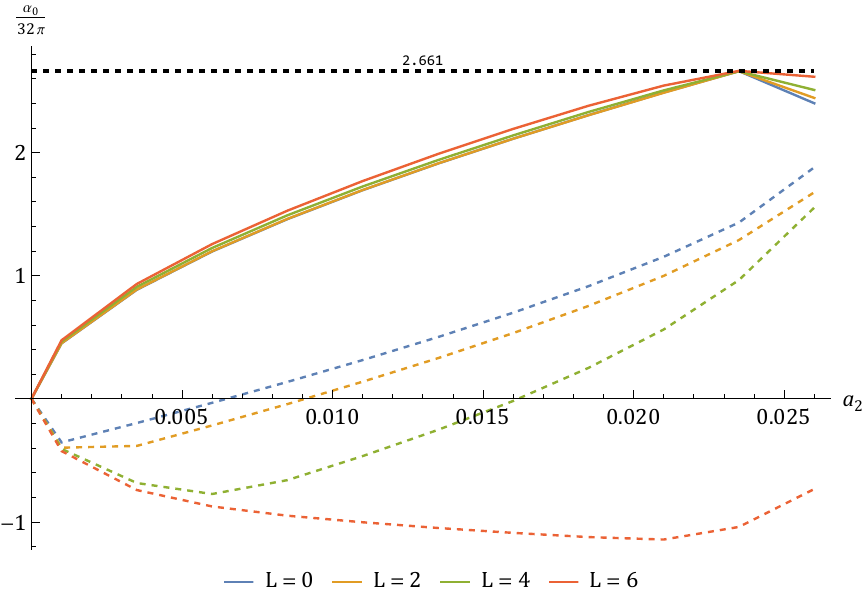}
     \end{subfigure}
     \begin{subfigure}{0.47\textwidth}
         \centering
         \includegraphics[width=1\textwidth]{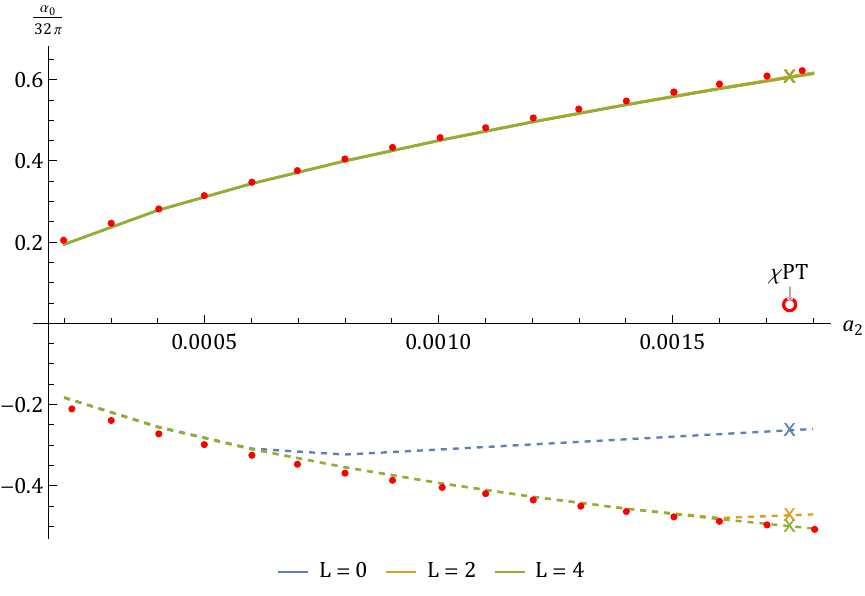}
     \end{subfigure}
      \caption{a) Upper and lower bounds on the quartic coupling vs. $a_2$ using the $F_B$-bootstrap. b) Zoomed-in version of the same plot near small $a_2$. The red points are taken from \cite{Lopez:1976zs}. The red circle is at $\lambda=0.046$ and is the value of the quartic coupling from 2-loop $\chi_{PT}$ at $a_2 = a_2^{(pion)} = 0.00175$.}
      \label{alpha0vsa2}
\end{figure}
Unlike the case when $a_2$ is not fixed, we observe that when $a_2$ is fixed and small, convergence is achieved with just the first few spins. This is explained by the fact that fixing scattering lengths imposes polynomial boundedness of the amplitude and leads to an effective truncation in spin. For instance, in the derivation of the Froissart bound (see e.g. \cite{yndurain, HS}), $a_2$ is held fixed, and the spin-sum entering in the definition of the forward scattering amplitude beyond $\ell>L$, with $L\sim \sqrt{s}\log s/s_0$ works out to be exponentially small---this immediately leads to the Froissart bound on setting $a_\ell(s)=1$ for $\ell<L$. Here $s_0$ is needed for dimensional grounds.  To see this more explicitly, we refer the reader to Appendix \ref{ScatteringLengths} where we have written the dispersive representation of the scattering lengths. As is evident, this involves an integral over the absorptive part of the amplitude. It can be checked that each partial wave spin contributes positively to the absorptive part. Thus, if the scattering lengths are fixed to be small and positive, the number of partial wave spins contributing also gets restricted, and we expect the bounds to converge with fewer spins. The numerics bear out this expectation.

\subsection{Minimizing $a_0$, keeping $a_2$ fixed}

\begin{figure}[hbt!]
\centering
\includegraphics[scale=0.7]{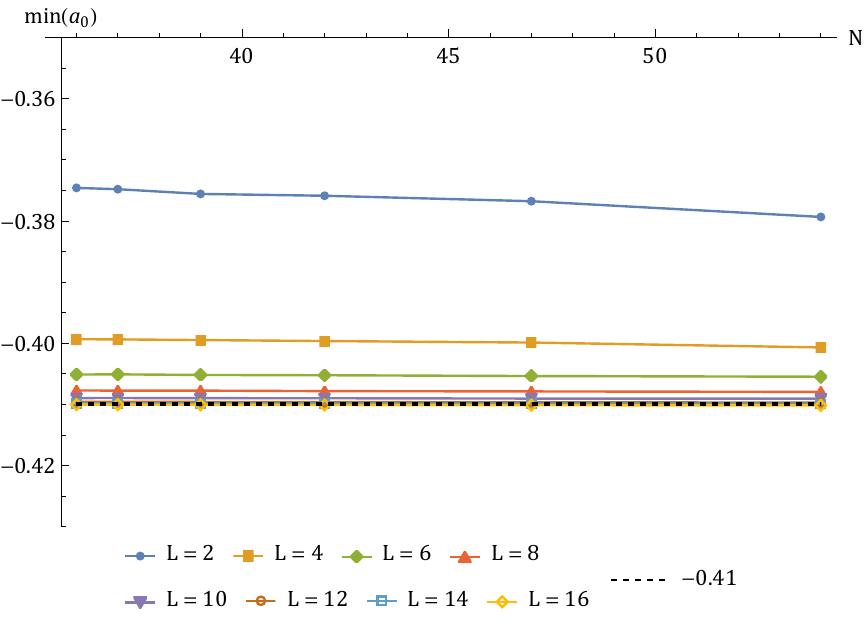}
\caption{Lower bound for $a_0$ using $D_B$-bootstrap when $a_2 = a_2^{(pion)}$. Unitarity was imposed up to $L=16$ and $s_{max}\approx 4950$. Null constraint $-10^{-12} \le \mathcal{W}_{-1,2} \le 0$ was imposed. }
\label{mina0}
\end{figure}

The scattering lengths are known to be unbounded from above \cite{Sboot2} but bounded from below. For the S-wave scattering length $a_0$, previous studies \cite{Sboot2, LopMen} quote a value for the lower bound around $-1.7$. We search for the minimum of $a_0$ when $a_2 = a_2^{(pion)}$ and find (see figure \ref{mina0})
\begin{equation}
    \min(a_0)|_{a_2 = a_2^{(pion)}} \approx - 0.41
\end{equation}

\section{$\rho_{KK}$ from the numerical bootstrap}
With the advent of the LHC, collision energies have increased by orders of magnitude. For example, proton-proton collisions are now accessible at energies $\sqrt{s} \sim 10$ TeV so that $\log(s/m^2) \sim O(10)$. In particular, experiments can now test the rigorous bounds derived using dispersion relations and, thereby, examine the validity of the QFT axioms that go into deriving them. In this section, we consider the observable $\rho_{KK}(s)$ introduced by Khuri and Kinoshita \cite{KK1}, which is defined for elastic processes via
\begin{equation}
    \label{RhoDef}
    \rho_{KK}(s) = \frac{\text{Re} \mathcal{M}(s,0)}{\text{Im} \mathcal{M}(s,0)}\,.
\end{equation}
In the case of 2-2 scattering of identical massive scalar particles, dispersion relations imply that at large energies, $\rho$ and the total cross section $\sigma$ in the forward limit are related by a simple relation \cite{Bronzan:1974jh,Menon:1998cm} given as \footnote{It is unclear, however, at what $s$-values this formula is a good approximation. For the modest $s$-values considered in this paper, we could not demonstrate this formula for all cases convincingly.}
\begin{equation} \label{RhoLargeS}
\lim_{s \rightarrow \infty} \rho_{KK}(s) = \frac{\pi }{2}  \frac{\partial\log \s (s)}{\partial \log s}\,.
\end{equation}
The above equation essentially means that the asymptotic behaviour of $\rho_{KK}(s)$ captures the asymptotic rate of growth of the total cross-section. For example, for amplitudes saturating the Froissart bound, i.e. $\sigma(s) \sim  \log^2 (s)$, the above formula gives $\rho_{KK}(s) \sim  1/ \log(s)$. Figure \ref{rhoExpt} shows the latest measurement of $\rho_{KK}$ for proton-proton scattering by the ATLAS detector. One can see from the figure that $\rho_{KK}$ changes sign from negative to positive and then turns around and starts to slope downwards at high energies. However, the energies probed currently at the LHC are not large enough to infer the exact large $s$ behaviour of $\rho$. So, a comparison with the dispersion relation prediction \eqref{RhoLargeS} is not possible. Our objective is to use the numerical bootstrap to study the behaviour of $\rho_{KK}$ at intermediate energies for which the experimental data exists.  In particular, we want to see if we can reproduce some of the features for $\rho_{KK}$ as observed in $pp$ experiments. Quite interestingly, we are able to construct some S-matrices that do display similar features, which we will discuss next. We note here the threshold behaviour for $\rho_{KK}$. In the case where there is a finite spin-0 scattering length $a_0$, we have
\be
\rho_{KK}(s)\sim \frac{2}{a_0\sqrt{s-4}}\,,
\ee
which means that $\rho_{KK}(s)\rightarrow \pm \infty$ depending on the sign of $a_0$.
If $a_0$ is infinite, then writing $Im f_0(s\rightarrow 4)=b_0$, we have
\be
\rho_{KK}(s)\sim \sqrt{\frac{1-b_0}{b_0}}\,.
\ee
This is the situation which gives rise to the maximum coupling. In that case, we find that $b_0\rightarrow 1$ and thus $\rho_{KK}\rightarrow 0$ at $s =4$. 

\subsection{$\rho_{KK}$: Minimising $\lambda$, keeping $a_0, a_2$ fixed}
\begin{figure}[H]
     \centering
     \begin{subfigure}{0.6\textwidth}
         \centering
         \includegraphics[width=1\textwidth]{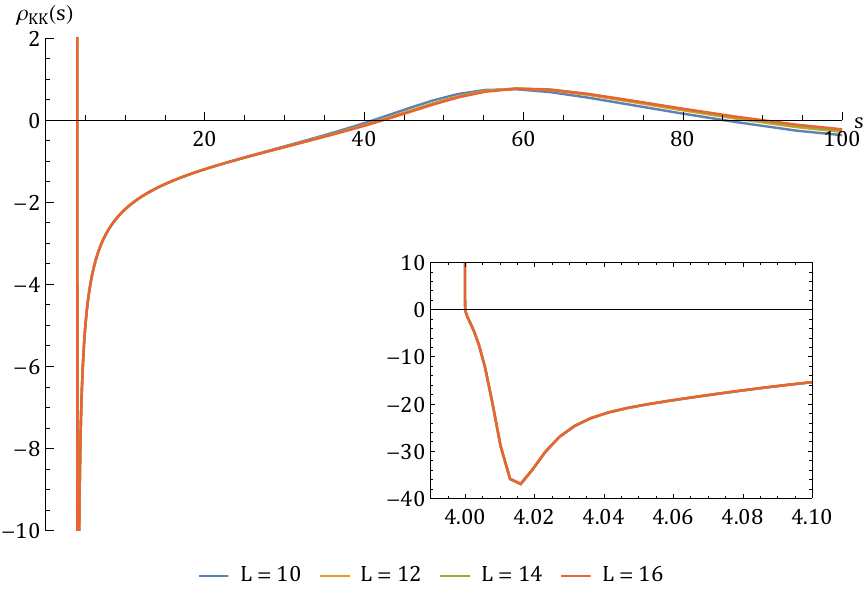}
         
         \caption{} \label{rhoa0a2fixedDyson}
     \end{subfigure}
     \begin{subfigure}{0.35\textwidth}
         \centering
         \includegraphics[width=1\textwidth]{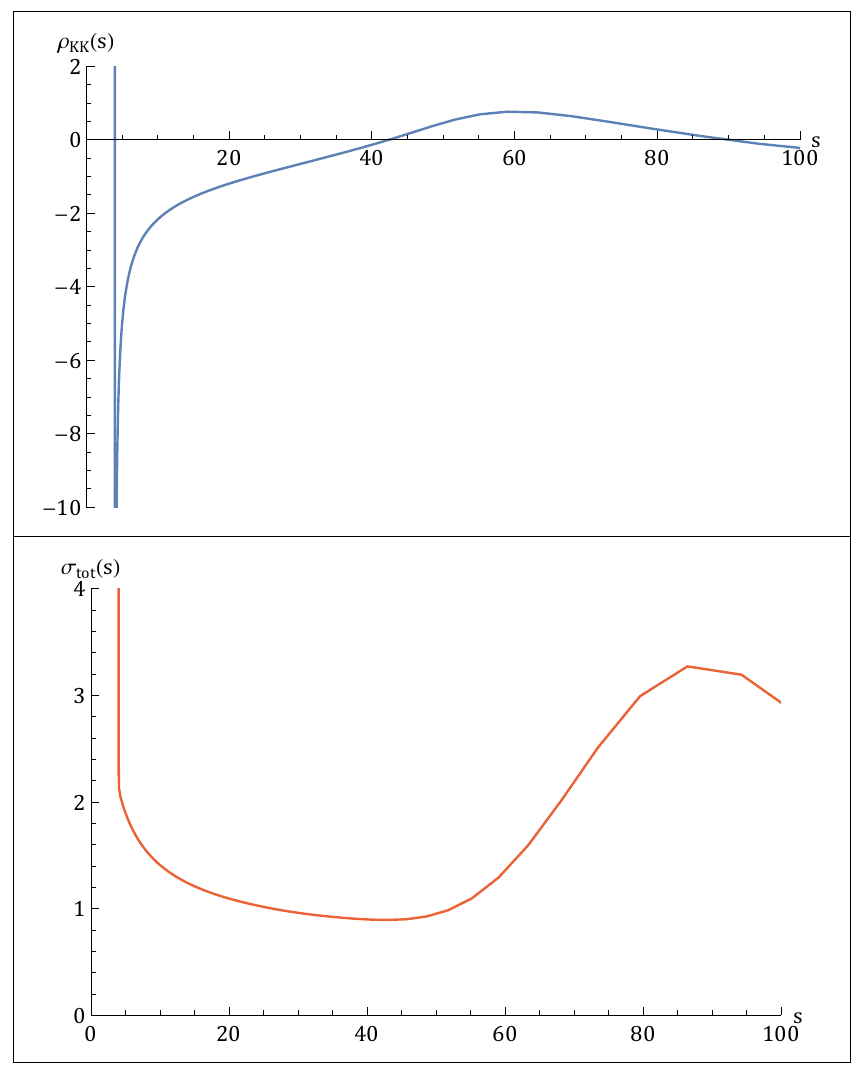}
         \caption{}
         \label{rhosigmaLmax16lambdaminDysona0a2fixed}
      \end{subfigure}
      \caption{Left: Behavior of $\rho_{KK}$ with $s$ for the amplitude with the minimum quartic coupling with $a_0=a_0^{(pion)}$ and $a_2=a_2^{(pion)}$. Right: Behavior of $\rho_{KK}$ and $\sigma_{tot}$ with $s$ for the same amplitude at $L=16$.}
      \label{rhoa0a2fixedlambdamin}
\end{figure}

In Appendix \ref{Comparison}, we study $\rho_{KK}$ for maximum coupling---see fig.(\ref{Inverserhocomparisonalpha0max}). There it is clear that no change of sign occurs in $\rho_{KK}$. In this section, we study $\rho_{KK}$ for the amplitude with the minimal quartic coupling when the spin 0 and spin 2 scattering lengths are fixed to the pion values, i.e. $a_0 = a_0^{(pion)}= 0.22$ and $a_2 = a_2^{(pion)}= 0.00175$ \cite{colangelo}. The numerics are done using the $D_B$-bootstrap with unitarity imposed up to $L =16$ and $s_{max} \approx 4950$. We also impose the null constraint $-10^{-12}\leq \mW_{-1,2}\leq 0$. Figure \eqref{rhoa0a2fixedlambdamin} shows our findings.

Near $s \approx 4$, $\rho_{KK}$ is positive as expected because $a_0$ is fixed to be positive. But then it immediately dives and becomes negative. Then we notice the first turnover and a change to positive values. We denote the location of this zero by $s_c^{(1)}$. After this, we see a peak. The zero $s_c^{(1)}$ and the peak are features, very similar to what is observed in the experimental $pp$ scattering plot (see \ref{rhoExpt}). However, quite interestingly, in our case, $\rho_{KK}$ becomes negative again after the peak. We denote this zero by $s_c^{(2)}$. In Appendix \ref{Comparison}, we show that $\rho_W$-bootstrap gives identical results. In Appendix \ref{LeafRhos}, we plot $\rho_{KK}$ for several other amplitudes taken from the leaf plot of \cite{Miro} and find that in many cases, there may even be more than one peak and more than two zeroes before $\rho_{KK}$ asymptotes. This indicates that $\rho_{KK}$ for $pp$ scattering might also undergo multiple changes in sign before asymptoting at large energies unlike what figure \ref{rhocartoon} suggests.

\subsubsection*{How does low-energy data affect the behavior of $\rho_{KK}$?}
\begin{figure}[H]
\centering
\includegraphics[scale=1.2]{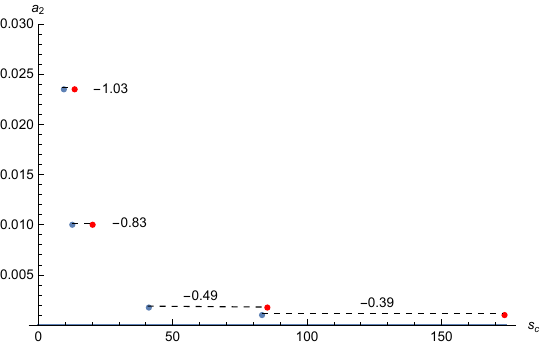}
\caption{Change in the difference between the two zeroes --- $s_c^{(1)}$ (blue dot) and $s_c^{(2)}$ (red dot) of $\rho_{KK}$ on the $s$-axis as we vary $a_2$. The corresponding values of $\lambda_{min}$ have been denoted adjacent to the plot points. The above plot was obtained using $D_{B}$-bootstrap with unitarity imposed up to $L=10$ and $s_{max}\approx 4950$.}
\label{a2crossing}
\end{figure}

We study how the location of zeroes $s_c^{(1)}$ and $s_c^{(2)}$  changes as we vary $a_0$ and $a_2$. We find that changing $a_0$ only changes the value of $\rho_{KK}$ near $s = 4$, leaving the location of the two zeros unchanged. However, changing $a_2$ has a significant effect on the location and the separation of the two zeroes, which we show in figure \ref{a2crossing}. This indicates that the behavior of $\rho_{KK}$ beyond the threshold is largely determined by the value of $a_2$. This observation is also supported by $\rho_{KK}$ for the amplitude we consider next. 

\subsection{$\rho_{KK}$: Minimising $a_0$, keeping $a_2$ fixed}
\begin{figure}[H]
     \centering
     \begin{subfigure}{0.6\textwidth}
         \centering
         \includegraphics[width=1\textwidth]{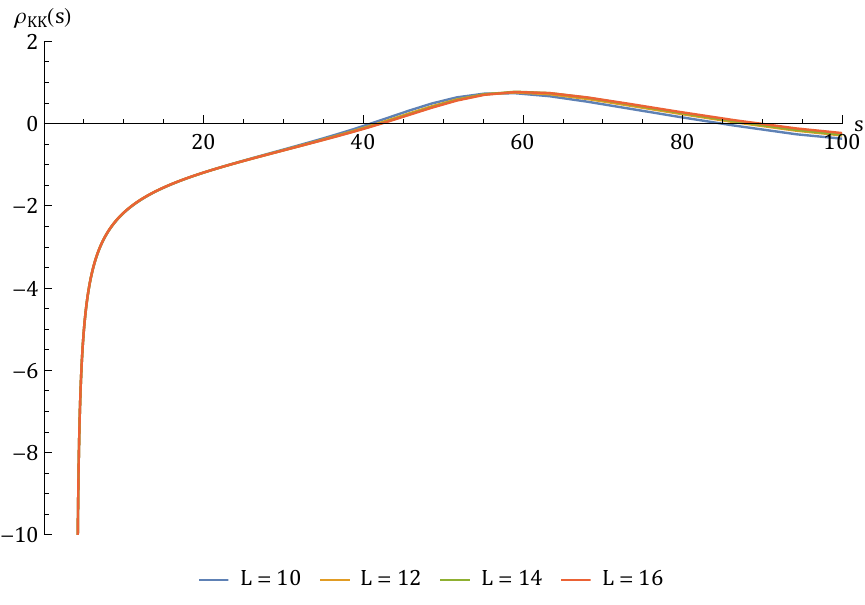}
         
         \caption{} \label{rhoa2fixedmina0Dyson}
     \end{subfigure}
     \begin{subfigure}{0.35\textwidth}
         \centering
         \includegraphics[width=1\textwidth]{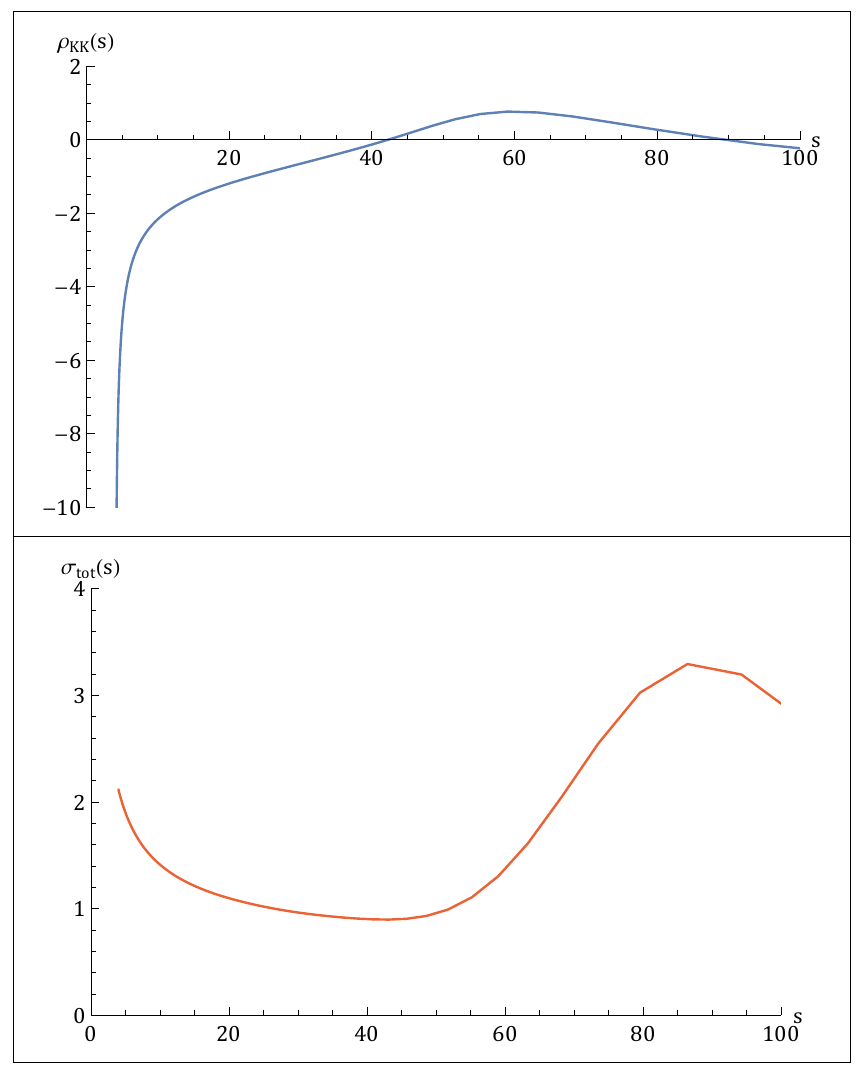}
         \caption{}
         \label{rhosigmaLmax16mina0Dysona2fixed}
      \end{subfigure}
      \caption{Left: Behavior of $\rho_{KK}$ with $s$ for the amplitude with the minimum $a_0$ and fixed $a_2 = a_2^{(pion)}$. Right: Behavior of $\rho_{KK}$ and $\sigma_{tot}$ with $s$ for the same amplitude at $L = 16$.}
      \label{a0minRho}
\end{figure}

The next amplitude we study $\rho_{KK}$ for is the one with the minimum $a_0$ when $a_2 = a_2^{(pion)}$. We studied this in section 5.2 and found $ \min(a_0)|_{a_2 = a_2^{(pion)}} \approx - 0.41$. Figure \ref{a0minRho} shows the behaviour of $\rho_{KK}$. The behaviour here is quite similar to what we found in fig. (\ref{rhoa0a2fixedDyson}). This is suggestive of the more important role that $a_2$ plays in determining the higher $s$ behaviour of $\rho_{KK}$.

\subsection{Phenomenological $\rho_{KK}(s)$ using bootstrap}
\begin{figure}[H]
     \centering
     \begin{subfigure}{0.47\textwidth}
         \centering
         \includegraphics[width=1\textwidth]{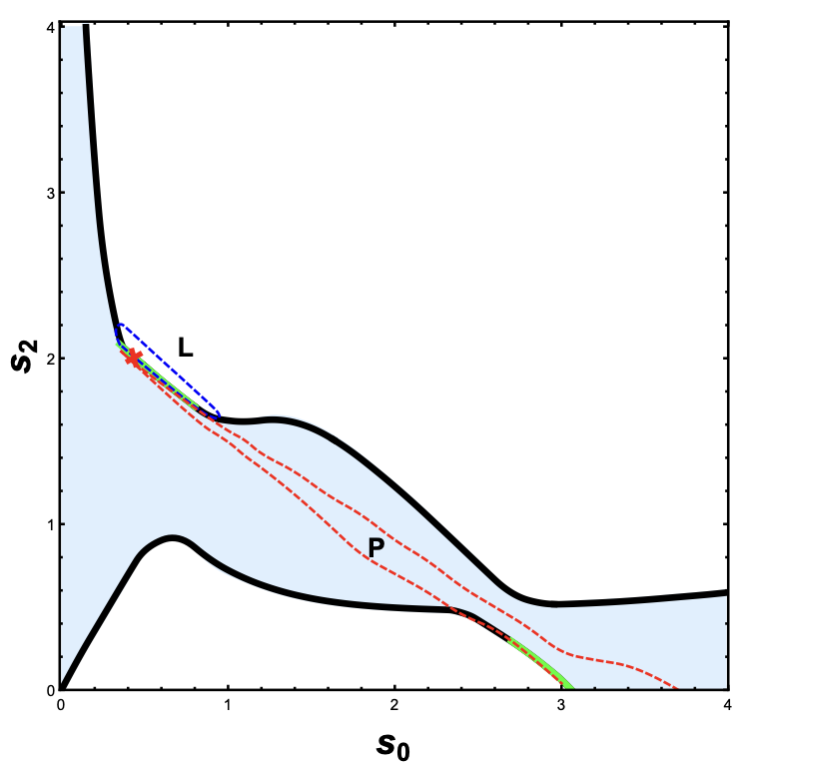}
      \caption{} \label{fig:river}
     \end{subfigure}
     \begin{subfigure}{0.47\textwidth}
         \centering
         \includegraphics[width=1.1\textwidth, height=8cm]{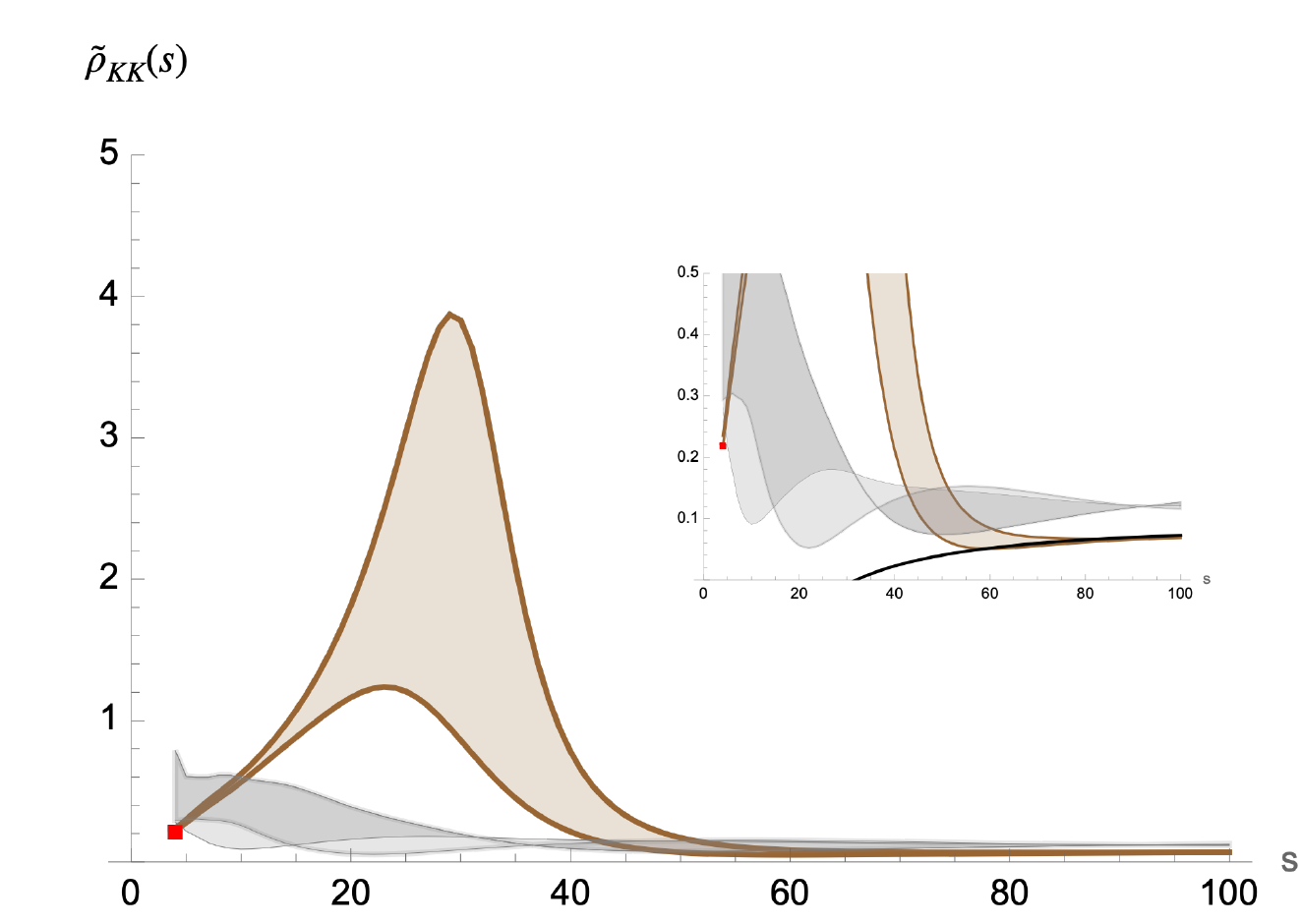}
     \caption{}  \label{fig:rhorho}
     \end{subfigure} 
      \caption{Left: The pion river obtained in \cite{ABAS}. $s_0, s_2$ are the isospin-0 and isospin-2 Adler zeros. The red cross is the $\chi-PT$ value. The L and P indicate the lake and peninsula regions obtained in \cite{andrea}. The green lines indicate where resonances reggeized. Right: The behaviour of $\tilde\rho_{KK}=2(s-4)^{-\frac{1}{2}}\rho_{KK}^{-1}(s)$ on the upper boundary. The brown band is for the S-matrices corresponding to $s_0=0.35$ to $s_0=0.4$, and all exhibit reggeized trajectories. The red box indicates experimental pion $a_0$ values. The black line in the inset is a fit using $\rho_{KK}=\pi/\log(s/s_0)$ with $s_0=32$.}
\end{figure}

\begin{figure}[H]
\centering
\includegraphics[scale=0.8]{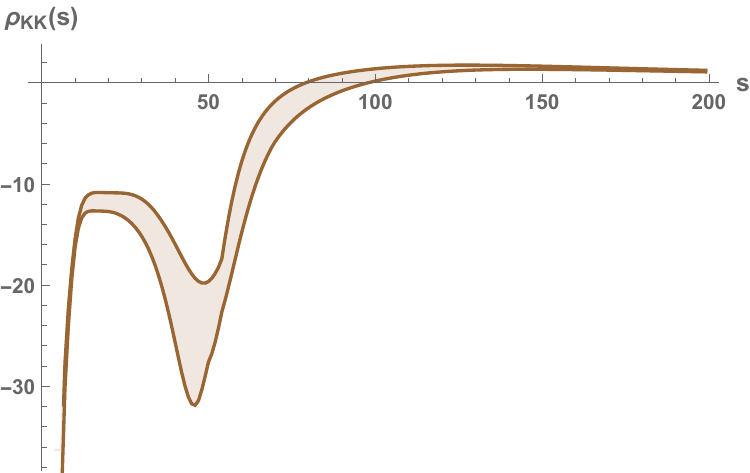}
\caption{Behavior of $\rho_{KK}(s)$ for the reaction $\pi^+ + \pi^+ \rightarrow \pi^+ + \pi^+$, plotted for S-matrices from $s_0=0.35$ to $s_0=0.4$ which exhibit approximately linear Regge trajectories.}
\label{pppp}
\end{figure}

In this section, we will briefly examine $\rho_{KK}$ in the context of the pion bootstrap\footnote{To repeat, we set $m_\pi=1$.}. In modern times, pion bootstrap was initiated in \cite{andrea} and examined in further detail in other contexts in \cite{BHSST, ABAS}. We will use the pion S-matrices obtained in \cite{ABAS} and the reader is referred to that paper for further details. The idea of the pion bootstrap was to use the S-matrix bootstrap in the context of 2-2 scattering of $\pi^0, \pi^\pm$ in QCD (ignoring electromagnetic interactions). The space of S-matrices was parametrized by the isospin-0 and isospin-2 Adler zeros. In \cite{andrea}, the spin-1 $\rho$ resonance (not to be confused with $\rho_{KK}$) was used as input. Then unitarity confined theories to lie {\it outside} a region dubbed as the ``lake" (L). As is clear from fig.(\ref{fig:river}), this gives a very big space of allowed S-matrices. Then putting in the known phenomenological scattering lengths as further input (importantly, $a_0 = a_0^{(pion)}$, a restricted region called ``peninsula" (P) was obtained. The QCD point is indicated by a red cross in Fig. (\ref{fig:river}). In \cite{BHSST, ABAS}, the dispersive constraints on the spin-2 scattering length and a similar constraint motivated by phenomenological observations on the spin-0 scattering length were imposed without imposing the phenomenological constraints directly. This gave rise to a bigger region, indicated in blue in the figure, dubbed as the ``river". The QCD-point (in the sense of the Adler-zeros) was strikingly close to a kink-type feature and remarkably gave very good scattering lengths. Furthermore, in \cite{ABAS}, it was observed that the peaks in the partial wave amplitudes lined up in small regions, indicated by green in the plot. Approximately the leading Regge trajectory, including the spin-0 resonance, obeyed:
\be \label{regg}
\alpha(t)=\alpha_0+\alpha' t\approx 0.38+0.51 m_\rho^{-2} t\,,
\ee
where $m_\rho$ is the real part of the rho-meson mass.
This is roughly what is expected from the Regge behaviour observed in experiments which gives $\alpha_0=0.27, \alpha'=0.54$. According to Regge theory and assuming an Eikonal model \cite{donnachielandshoff}, eq.(\ref{regg}) leads to the total forward scattering cross-section behaving like $\sigma_{tot}\propto \log^2 s/s_0$, where $s_0$ is an undetermined constant. This is the expected behaviour interpolated from eq.(\ref{regg}), which can be extracted from low-lying spin resonances to asymptotic energies. In practice, we cannot observe this asymptotic behaviour directly from the bootstrap. The green region in the lower boundary of the river had large scattering lengths as well as effective ranges and did not agree with phenomenology. 

In fig.(\ref{fig:rhorho}), we show the behaviour of $\tilde\rho_{KK}(s)=2 (s-4)^{-1/2}/\rho_{KK}(s)$. This quantity directly gives the scattering length at $s=4$. Quite strikingly, there is a distinct change in behaviour in $\rho_{KK}$ for the S-matrices close to the QCD-point! A similar observation is made for the Reggeized S-matrices for the lower boundary of the river, except that these gave large scattering lengths. 
From eq.(\ref{expKK}) we expect that for large energies, $\tilde\rho_{KK}=2 /(\pi s^{1/2}) \log(s/s_0)$. Fitting this form beyond $s\sim 70$ to the data in fig.(\ref{fig:rhorho}), gives $s_0\approx 32$. This is very interesting since \cite{martinroy} quotes the phenomenological value of $s_0=17$ in this case to fit the total scattering cross-section at $s=50$. 

The following points are noteworthy:
\begin{itemize}
\item The $\rho_{KK}$ for $\pi^0\pi^0$  is distinct from the one in $pp$ scattering. Since we are plotting inverse $\rho_{KK}$ in fig.(\ref{fig:rhorho}), it is clear that here the quantity has the same sign throughout since if it had a zero, it would show up as a pole in $\tilde\rho_{KK}$. The difference arises since the strong force in $pp$ scattering at low energies gives rise to an attractive potential and hence a negative scattering length. Since $pp$ cross-section is expected to rise, $\rho_{KK}$ is expected to approach zero from above. Khuri-Kinoshita pointed out that it should change sign somewhere, which is what the experimental data bears out.
\item For the S-matrices lying beyond the green ``Regge" region in fig.(\ref{fig:river}), eventually $\rho_{KK}$ does cross the real axis {\it from above}. All the S-matrices in the lower river boundary have this feature, except the ones in the green ``Regge region".
\item We  plot $\rho_{KK}(s)$ for the reaction $\pi^+ + \pi^+ \rightarrow \pi^+ + \pi^+$ for the ``Reggeized" S-matrices in fig.(\ref{pppp}). This reaction is closer to the $pp$ scattering carried out in experiments. Note that there is indeed a zero, and the $\rho_{KK}(s)$ goes to zero from above, much like Khuri-Kinoshita pointed out. This behaviour is only true for S-matrices close to the ``Regge" region, providing another argument for the QCD-like point being in its vicinity.
\end{itemize}

\subsection{$d > 4$ unitarity at high energies: A preliminary attempt}

\begin{figure}[H]
\centering
\includegraphics[scale=0.8]{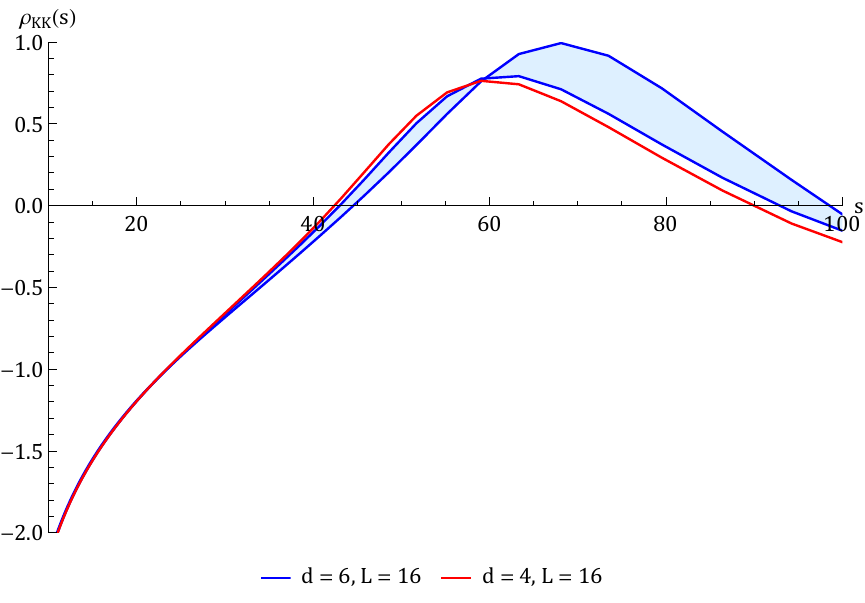}
\caption{Behaviour of $\rho_{KK}(s)$ for the amplitude with minimum quartic coupling and $a_0$, $a_2$ fixed to the pion values after imposing higher $d = 6$ unitarity constraints. The red line denotes the usual $\rho_{KK}(s)$ obtained by using $d = 4$ unitarity constraints at $L = 16$. The shaded blue band denotes the $\rho_{KK}(s)$ obtained by using $d = 6$ unitarity constraints at $L = 16$. There are two blue lines. The blue line with a bigger peak and a slower fall-off is when $s_0\approx50$, whereas the blue line that is closer to the red line is for $s_0\approx450$. We use $D_B$-bootstrap and impose unitarity till $s_{max} \approx 4950$.} 
\label{rho4d6ds052and448}
\end{figure}

If extra spatial dimensions exist, the physics at sufficiently high energies should be sensitive to them. The energy scale beyond which that happens gives a measure of the characteristic length of the extra dimensions. Recently \cite{maharana} established the existence of dispersion relations in the presence of Kaluza-Klein (KK) modes. In scattering experiments, the signature would be the production of KK particles at very high energies. Experiments conducted so far at the LHC have found no signs of extra spatial dimensions, giving us at least a lower bound for the energy scale at which they become important. We would like to model this using the bootstrap. We fix a scale $s_0$ below which we impose $(3+1)~d$  unitarity and fix the spin 0 and spin 2 scattering lengths, $a_0$ and $a_2$, to pion values. Beyond $s_0$, we impose higher dimensional unitarity. The partial waves in general $d$ are given by
\begin{equation}
\label{fellHigherd}
    f_{\ell}(s) = \frac{\Gamma\left( \frac{d-3}{2}\right) \ell!}{2^{d+1}\pi^{\frac{d-1}{2}}\Gamma\left(\ell+d-3\right)} \frac{(s-4)^{\frac{d-3}{2}}}{s^{\frac{1}{2}}} \int_{-1}^{1} \frac{dz}{2}(1-z^2)^{\frac{d-4}{2}}\mathcal{C}^{\frac{d-3}{2}}_{\ell}(z) \mathcal{M}(s,t)
\end{equation}
Here, $\mathcal{C}^{\frac{d-3}{2}}_{\ell}(z)$ are the $d$-dimensional Gegenbauer polynomials. In our conventions, the unitarity constraint is given by the same equation \eqref{eq:uni} in all dimensions. We then study $\rho_{KK}$ for the amplitude with the minimum quartic coupling and see how the low energy ($s < s_0$) behaviour of $\rho_{KK}$ changes compared to the case where no extra dimensions appear. The choice of $s_0$ is important. In figure \ref{rhoa0a2fixedlambdamin}, the crossover from negative to positive and then a peak are two features that the experimental $\rho_{KK}$ data for $pp$ scattering also shows (figure \eqref{rhoExpt}). So we will pick $s_0$ to be near and beyond the peak and study the effect on the location of the zero and the peak due to extra dimensions. 

Fig.(\ref{rho4d6ds052and448}) reports our findings. As is evident from the figure, imposing higher dimensional unitarity beyond the peak (i) moves the peak to the right (ii) moves the second zero to the right. As we increase $s_0$, the result begins to coincide more with the $d=4$ case, as we would naively expect. Thus extra dimensions may not be the explanation for the faster fall-off in fig.(\ref{rhoExpt}). A likely explanation of the deviation is the non-applicability of the high-energy asymptotic behaviour to get the dispersive band, as reviewed in \cite{Matthiae:1994uw}. It has also been proposed that the faster drop could be due to a new trajectory called the ``odderon" \cite{TOTEM:2019}\footnote{The leading particle on this trajectory is a 3-gluon bound state.}. It will be interesting to explore this further using the bootstrap.

\section{Discussion}
In this paper, we introduced a new basis to implement numerical S-matrix bootstrap using the CSDR/LCSDR. We compared with existing methods and found very good agreement. We also found that there is improvement in convergence using our approach. There are several key findings which we summarize below and 
discuss some open directions. 

One of the main lines of investigation in this paper was to examine $\rho_{KK}(s)$. Different high-energy behaviours lead to different predictions for this quantity, relying on a dispersion relation. One main feature is the fact that it can change sign from negative to positive, which is what is observed in $pp$-scattering experiments. This sign change is something that we observed in our case studies as well. In particular, we found that the $s$-value at the crossing was sensitive to the spin-2 scattering length but less dependent on the spin-0 scattering length. Another point that we observed was that to get convergent S-matrices even at modest values of $s$, we needed to impose unitarity up to very high energies. This was a feature that was confirmed by both the $\rho_W$-bootstrap as well as the $D_B$-bootstrap. To be specific, as fig.(\ref{rhoa0a2fixedconvergencewithsmax}) shows, imposing unitarity up to different maximum $s$-values alters the location of the second zero in $\rho_{KK}$. 

In fig.(\ref{rhocartoon}), we had indicated several possible behaviours of $\rho_{KK}$ \cite{Matthiae:1994uw}. From the bootstrap, we see features in common but also differences. For instance, we see the change of sign, followed by a peak. But the main (and possibly important) difference is that we also witness a second change of sign. This feature is corroborated by both $D_B$-bootstrap as well as $\rho_W$-bootstrap. This suggests that unlike fig.(\ref{rhocartoon}), there could be multiple changes of sign before $\rho_{KK}$ asymptotes at high energies. In fact, the TOTEM and ATLAS values in fig.(\ref{rhoExpt}) do indicate a sharper fall-off, possibly hinting at a second zero. 

This raises the question: How do we know that we have reached asymptotic high energies? We do not have a clear answer to this question\footnote{It is possible that subleading terms in the Froissart-Martin bound are important to retain in order to explain the data--see \cite{rohinime} for example.}. In the literature \cite{Bronzan:1974jh,Menon:1998cm}, for asymptotic high energies, several papers have looked at what is called the quasi-local differential form, which leads to equations like eq.(\ref{RhoLargeS}). This is a remarkable form since, instead of the integration that one faces in the dispersion relation, one gets a simple local relation between the real and imaginary parts through such a relation. In fig.(\ref{rhosigmaLmax16lambdaminDysona0a2fixed}), the zeros of $\rho_{KK}$ do appear to line up with the turning points of $\sigma_{tot}$. Nevertheless, we did not find any conclusive evidence that a local relation like eq.(\ref{RhoLargeS}) holds at the intermediate values of energies that we have investigated in this paper---for fig.(\ref{rhosigmaLmax16lambdaminDysona0a2fixed}), multiplying the RHS of eq.(\ref{RhoLargeS}) by a phenomenological factor of 0.2 appears to give a good fit between $s=32-250$ but not beyond. We believe it is important to understand and re-examine the situation with this quasi-local form in the future as it would be a very convenient way to impose high energy constraints, which can potentially improve convergence.

Another line of investigation would be to see if the numerics can be made more efficient by using other bounded functions like $\tanh (s)$ and its powers in the $D_B$-bootstrap. Such bounded functions are regularly used in the context of neural networks and it would be fascinating to use such techniques to further understand $\rho_{KK}$, similar to the recent attempt to understand the phase of the amplitude constrained by elastic unitarity in \cite{sashaneural}. It may also be worthwhile to use other standard wavelets which are used in wavelet transform in the $D_B$-bootstrap. Similar techniques will also be useful to analyse scattering in AdS space \cite{chuotAdS, vanRees:2022zmr}.

The most important line to investigate in the near future is the dual bootstrap problem, as it leads to completely rigorous bounds. As discussed in section 2, the Virasoro-Shapiro amplitude provides evidence that the CSDR may have a bigger domain of convergence than a fixed-$t$ dispersion relation, and the LCSDR may have the biggest domain of convergence among all three. This makes setting up the numerical bootstrap using CSDR/LCSDR more advantageous. This is because the dual approach involves constructing a dual Lagrangian, which consists of an integral over the domain of analyticity of the amplitude. The bigger the domain of analyticity, the more constraints are being imposed and, therefore, the stronger the bound.

\section*{Acknowledgements}
We thank Agnese Bissi, Andrea Guerrieri, Joan Miro and Chaoming Song for useful discussions on related topics considered in this paper. We thank Mehmet A. Gumus and Joan Miro for sharing their S-matrix data. We would also like to thank ICTS, Bangalore for allowing us to use their powerful clusters (Mario and Tetris) to perform some time-taking computations. AS thanks the participants of S-matrix Bootstrap V at Les Diablerets for illuminating discussions. 
AS acknowledges support from SERB core grant CRG/2021/000873. AZ has received support from the European Research Council, grant agreement n. 101039756.

\appendix
\section{Details of numerics}
\label{NumDetails}
We impose the unitarity condition in terms of the partial waves by imposing the positive semi-definiteness condition given in \eqref{eq:uni}. We repeat it here for convenience
\be
\begin{pmatrix}
1 + 2\text{Re}f_\ell(s) & 1 - 2\text{Im}f_\ell(s) \\
1 - 2\text{Im}f_\ell(s) & 1 - 2\text{Re}f_\ell(s) 
\end{pmatrix} \succeq 0, \quad s \ge 4\,.
\ee
where $\text{Im}f_\ell(s)$ is parametrized by the ansatz given in \eqref{ansatz}. To compute $\text{Re}f_\ell(s)$, we use the Roy equations as discussed in section 3. We impose unitarity for all spins up to some cut-off $\ell \le L$ and for a grid of $s$ values in the interval $(4, \infty)$. As is typical, we use a Chebyshev grid defined by 
\begin{equation}
    s(j) = \frac{\frac{4}{3} (1- \rho(j))^2+ 16 \rho(j)}{(1 + \rho(j))^2}, \quad \rho(j) =  e^{i \frac{j \pi}{203}}
\end{equation}
For all the bounds, while working with the LCSDR, we impose unitarity up to $s(j =190) \approx 265$, while with the CSDR, we go up to $s(j = 200) \approx 4950 $. We use the same grid to discretize the ``wavelet" parameter $\kappa \equiv \kappa (q)$. As described in the main text, this parameter controls the location of the peaks in our ansatz. The peaks are meant to capture the local features of the partial waves, and therefore, we sprinkle them along the same interval where we impose unitarity, i.e. $\kappa \in (4, \infty)$. We fix the parameter $\G$ that controls the width of the peaks to a constant $\approx 136$. This is a choice; we observe that any value of $\G$ which is not too small or too large leads to well-convergent numerics. We divide the basis elements into two sets - the ``low energy" wavelets that are placed from $q=1$ to $q = Q_1$ in steps of $qs_1$ and the ``high energy" wavelets that are placed from $q = Q_1+1$ to $q = Q$ in steps of $qs_2$. To claim convergence for any bound, we increase the number of parameters by decreasing $qs_1$ till we get a stable plateau of constant values for the bound for a large range of parameters. For the $D_B$-bootstrap, we fix $Q_1 = 177$, $Q =200$ and $qs_2 =1$, while for the $F_B$-bootstrap, we fix $Q =190$, $qs_2 =1$ and need to vary $Q_1$ for different $L$. For example, for the $\text{max}\left(\frac{\a_0}{32\pi}\right)$ plot using $F_B$-bootstrap (figure \ref{max_con}), we use $Q_1(L=0) = 180$, $Q_1(L=2) = 165$ and $Q_1(L=4) = 150$. The number of parameters used per spin (without $\a_0$ and $b_0$) is given by $N = \left\lfloor\frac{Q_1-1}{qs_1}\right\rfloor+ \left\lfloor\frac{Q-Q_1-1}{qs_2}\right\rfloor +2$. In addition to these, we have the extra parameters $b_0$ and the quartic coupling $\alpha_0$, making the total number of parameters in our ansatz $N_{tot} = N\left(\frac{L}{2}+1\right)+2$.   \\
The positive semi-definiteness condition is imposed using SDPB. The parameters we use for SDPB are summarized in table \eqref{SDPBTable}.
\begin{table}[hbt!]
\begin{center}
\begin{tabular}{ |c|c|c| } 
 \hline
 SDPB precision & 1024 (binary) \\ 
 \hline
 dualityGapThreshold & $10^{-7}$  \\ 
 \hline
 Mathematica internal precision & 200 (decimal) \\ 
 \hline
\end{tabular}
 \caption{SDPB parameters}
 \label{SDPBTable}
\end{center}
\end{table}

\begin{figure}[H]
\centering
\includegraphics[scale=0.8]{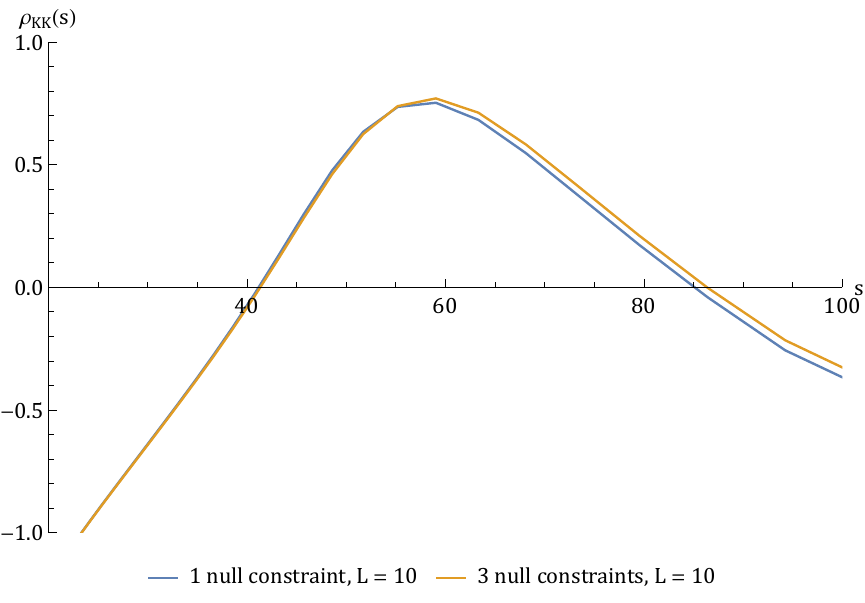}
\caption{Behaviour of $\rho_{KK}(s)$ for the amplitude with minimum quartic coupling and $a_0$, $a_2$ fixed to the pion values after imposing 1 null constraint and 3 null constraints. We use $D_B$-bootstrap and unitarity is imposed till $L=10$ and $s_{max} \approx 4950$.} 
\label{rhoa0a2fixedlambdamin1null3nulls}
\end{figure}

Additionally, while doing numerics using the $D_B$-bootstrap, we impose one null constraint, namely $-10^{-12}\leq \mW_{-1,2}\leq 0$, on the partial wave coefficients using the formula \eqref{WCDef}. We observe that our results don't change much by imposing additional null constraints. For example, figure \ref{rhoa0a2fixedlambdamin1null3nulls} shows the change in $\rho_{KK}(s)$ for the amplitude with minimum quartic coupling, $a_0 = a_0^{(pion)}$ and $a_2= a_2^{(pion)}$, when three null constraints are imposed instead of one. While imposing three null constraints, we impose $-10^{-12}\leq \mW_{-2,3}\leq 0$ and  $-10^{-12}\leq \mW_{-3,4}\leq 0$ in addition to imposing $-10^{-12}\leq \mW_{-1,2}\leq 0$. 

\begin{figure}[H]
     \centering
         \includegraphics[width=0.65\textwidth]{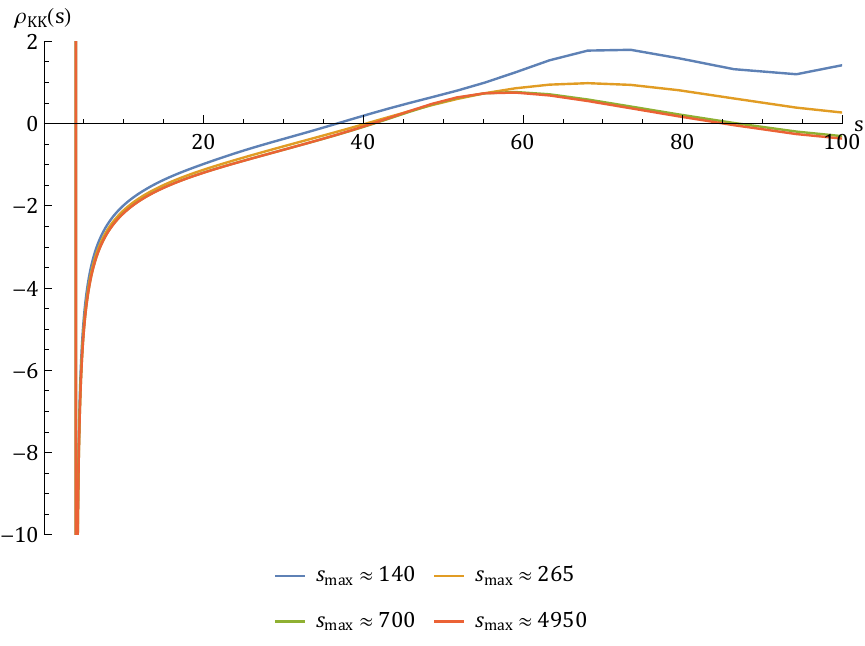}

      \caption{Convergence of $\rho_{KK}(s)$  with $s_{max}$ for the amplitude with minimum quartic coupling and $a_0$ and $a_2$ fixed to the pion values. We use $D_B$-bootstrap and impose unitarity till $L=10$.}  \label{rhoa0a2fixedconvergencewithsmax}
\end{figure}

We also verify convergence in $s_{max}$. Figure \ref{rhoa0a2fixedconvergencewithsmax} shows that $\rho_{KK}$ for the amplitude with minimum quartic coupling, $a_0 = a_0^{(pion)}$ and $a_2= a_2^{(pion)}$ doesn't change after we impose unitarity beyond a certain $s_{max}$.

\section{Representation for $a_0, a_2$ using CSDR}
\label{ScatteringLengths}
To obtain the formula for $a_0$ from CSDR (equation \eqref{disper0}), we first evaluate the integral at $s=4$. Note that at $s = 4$, the kernel $H_0 =\frac{96}{9 \tau^2-12 \tau -32}$ and $s_2^{(+)}|_{s=4}=\frac{1}{2} \tau \left(\sqrt{1-\frac{32}{9 \tau +8}}-1\right)$. Now, since $a_0=\frac{1}{32\pi}\lim_{s\to 4}\frac{1}{2} \int_{-1}^{1}\mathcal{M}(s,(s-4)(x-1)/2)dx$, we can substitute the expressions for $H_0$ and $s_2^{(+)}$ at $s=4$ to obtain the formula for $a_0$ from CSDR.
\be\label{eql0}
a _0=\frac{\alpha _0}{32\pi}+\frac{1}{32\pi}\int_{\frac{8}{3}}^{\infty}d\t\frac{96 \mathcal{A}_0\left(\tau ,\frac{1}{2} \tau  \left(\sqrt{1-\frac{32}{9 \tau +8}}-1\right)\right)}{\pi  \tau  \left(9 \tau ^2-12 \tau -32\right)} \,.
\ee
Similarly, we have the formula for $a_0$ from LCSDR.
\be
a_0=\frac{\alpha _0}{32 \pi }+\int_{\frac{8}{3}}^{\infty}d\t\frac{\left(-\frac{8}{3 \tau +4}+\frac{8}{3 \tau -8}+2\right) \mathcal{A}_0\left(\tau ,\frac{1}{18} \left(\sqrt{81-\frac{1536}{\tau ^3}}-9\right) \tau \right)-2 \mathcal{A}_0(\tau ,0)}{32 \pi ^2 \tau } \,.
\ee
An interesting feature of both formulas is that they both imply the inequality
\be\label{inta0eq}
a_0\geq \frac{\a_0}{32 \pi}
\ee
This inequality holds because the integral can be shown to be positive.

The formula for $a_2$ using the CSDR (although we use the formula \eqref{lam2scatt} for numerics) is given by
\be
a_2=\int_{\frac{8}{3}}^{\infty}d\t \frac{9 \mathcal{A}\left(\t, \frac{1}{2} \tau  \left(\sqrt{1-\frac{32}{9 \tau +8}}-1\right)\right)}{160 \pi ^2 (3 \tau +4)^3}+\frac{9 \sqrt{3} \tau  \mathcal{A}'\left(\t, \frac{1}{2} \tau  \left(\sqrt{1-\frac{32}{9 \tau +8}}-1\right)\right)}{5 \pi ^2 (3 \tau -8)^{3/2} (3 \tau +4) (9 \tau +8)^{3/2}} \,.
\ee
where $\mathcal{A}'(\t, s_2) = \frac{\partial}{\partial \t}\mathcal{A}(\t, s_2)$.

\section{String theory redux}
\label{StringRedux}
\begin{figure}[hbt!]
\centering
\includegraphics[width=\textwidth]{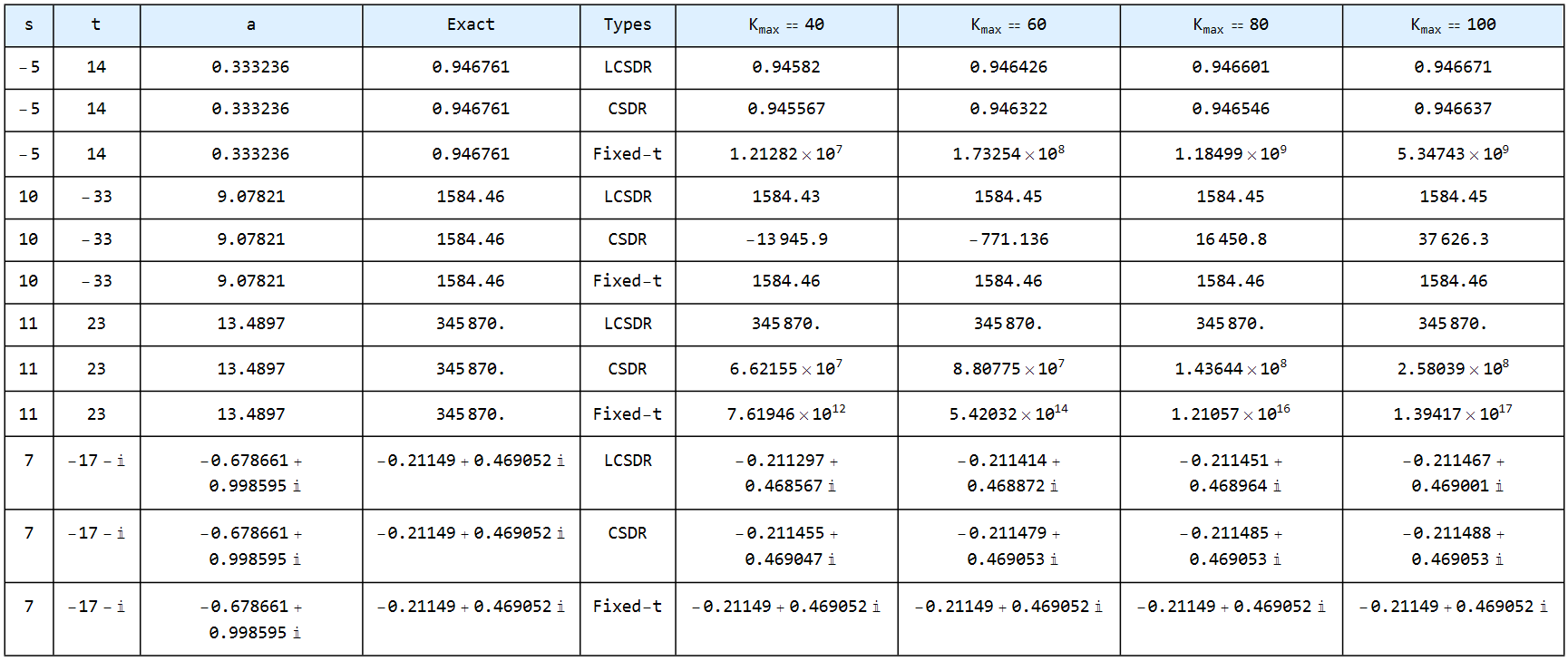}
\caption{Table for comparison of various dispersion relations.}
\label{tab:VS}
\end{figure}

A numerical comparison between the mass-level expansion for the tachyon amplitude in the main text 
obtained using fixed-$t$, CSDR and LCSDR is given in the table \eqref{tab:VS}.
The LCSDR always gives better convergence for each case.

\begin{figure}[H]
\centering
\includegraphics[width=0.65\textwidth]{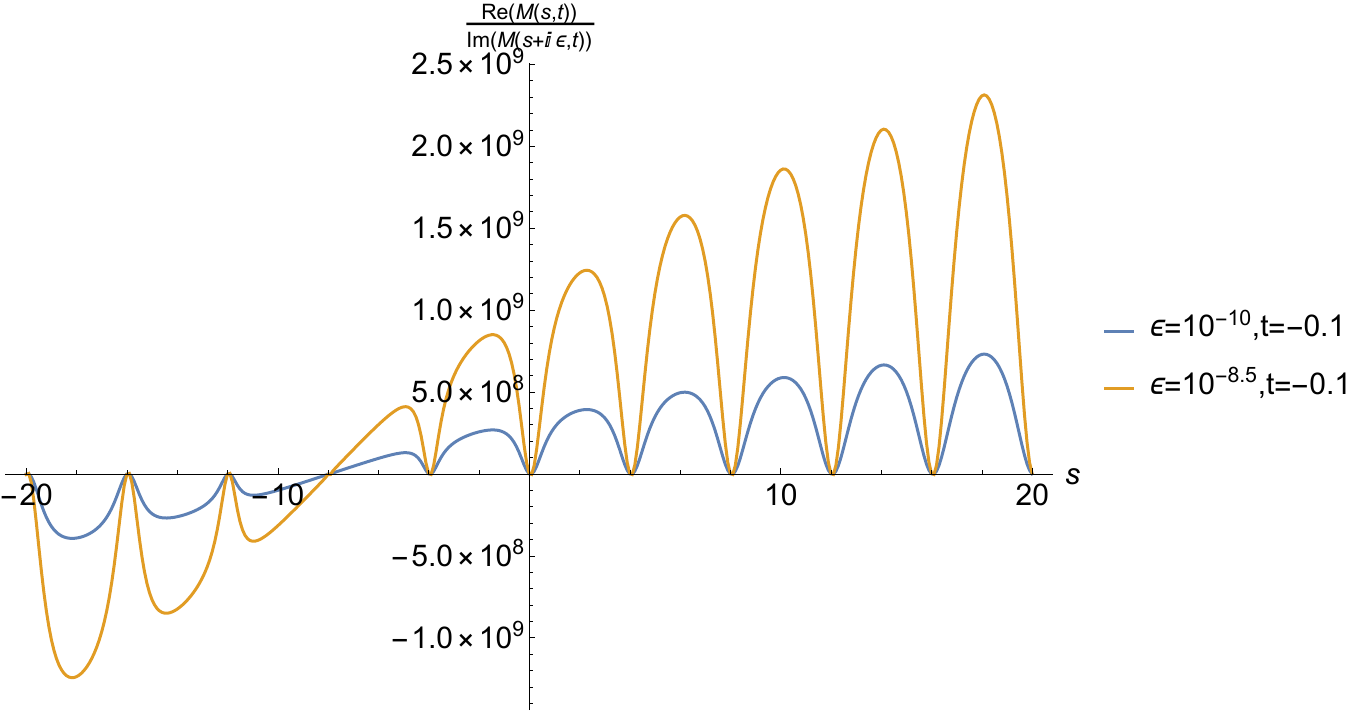}
\caption{$\rho_{KK}(s)$ for string theory.}
\label{rhostring}
\end{figure}

In the string theory case, $\rho_{KK}$ behaves as in figure \ref{rhostring}. Since the absorptive part has delta function support, to obtain this figure, we are forced to put an $i \epsilon$ regulator and also forced to work away from the strict forward ($t=0$) limit. This can be thought of as giving a finite width to the massive states. The oscillatory behaviour is expected. Contrary to the fall off predicted by Khuri-Kinoshita, the enveloping function is in fact an increasing function. This is indicative of the lack of polynomial boundedness of tree-level string theory. It will be fascinating to revisit $\rho_{KK}$ using one-loop string theory results.

\section{Comparison with existing methods}
\label{Comparison}
In this section, we shall compare our results with the numerical bootstrap set up in \cite{Miro}. They work with the following ansatz
\beq
\begin{split}
&\mathcal{M}_{\rho_W}(s,t,u)\:=\:\tau_0\:+\:\sum_{\textit{w}\:\in\: \mathcal{W}}\:\tau_{\textit{w}}\:\left(\sigma_s\left(\textit{w}\right)+\sigma_t\left(\textit{w}\right)+\sigma_u\left(\textit{w}\right)\right)\:\\&+\:\sum_{(\textit{w}_1,\textit{w}_2)\:\in\:\mathcal{W}^2}\:\tau_{\textit{w}_1,\textit{w}_2}\:\left[\left(\sigma_s\left(\textit{w}_1\right)\sigma_t\left(\textit{w}_2\right)+\sigma_s\left(\textit{w}_2\right)\sigma_t\left(\textit{w}_1\right)\right)+ \left( s\leftrightarrow u\right) + \left(t \leftrightarrow u\right) \right]
\end{split}
\eeq
where $\mathcal{W}$ and $\mathcal{W}^2$ are the wavelet grids parameterized by the number $N_{max}$, the details of which can be found in the appendix of \cite{Miro}. $\sigma_s(\textit{w})$ is the usual conformal mapping,
\beq
\sigma_s(\textit{w})\:=\:{\sqrt{\textit{w}-s}-\sqrt{4-s} \over \sqrt{\textit{w}-s}+\sqrt{4-s}}
\eeq
We calculate the partial waves $f_{\ell}(s)$ by integrating over $z$ \eqref{PWE} and impose the partial wave unitarity for $s\in \mathcal{W}$, up to some chosen $L_{max}$. We choose the grid values in $\mathcal{W}$ in the same way as \cite{Miro}, but we only consider points satisfying $s \leq s_{max}$. We also impose the improved positivity constraints as they do, given as
\beq
\mathcal{A}_{\rho_W}(s,t)\:-\:32\pi\:\sqrt{{s\over s-4}}\:\sum_{\ell\:=\:0}^{L_{max}}\:(2 \ell +1)\:\text{Im}f_{\ell}(s)\:P_{\ell}\left(1+{2 t \over s-4}\right)\:\geq\:0
\eeq
for $s\in \mathcal{W}$ and for $0\leq t \leq 4$. We call this the $\rho_W$-bootstrap approach.
Note that we do not include the high-energy improvement terms in this analysis.\par
\begin{figure}[H]
\centering
\includegraphics[width=\textwidth]{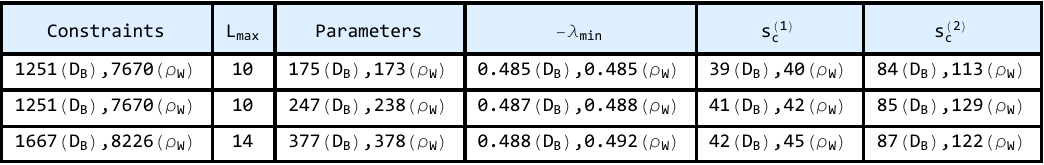}
\caption{Comparison of values obtained with the $D_B$ approach and the $\rho_W$ approach of \cite{Miro}. We compute $\lambda_{min}$ with $a_0$ and $a_2$ fixed to the pion values. The first two rows are made with $L = L_{max}\:=\:10$ and the last row is with $L = L_{max}\:=\:14$. $s_c^{(1)}$ and $s_c^{(2)}$ stand for the first and second zeros of $\rho_{KK}(s)$. The large number of constraints in the $\rho_{W}$ approach is due to the improved positivity constraints.}
\label{Tabcomp}
\end{figure}
\begin{figure}[H]
     \centering
     \begin{subfigure}{0.49\textwidth}
         \centering
         \includegraphics[width=1\textwidth]{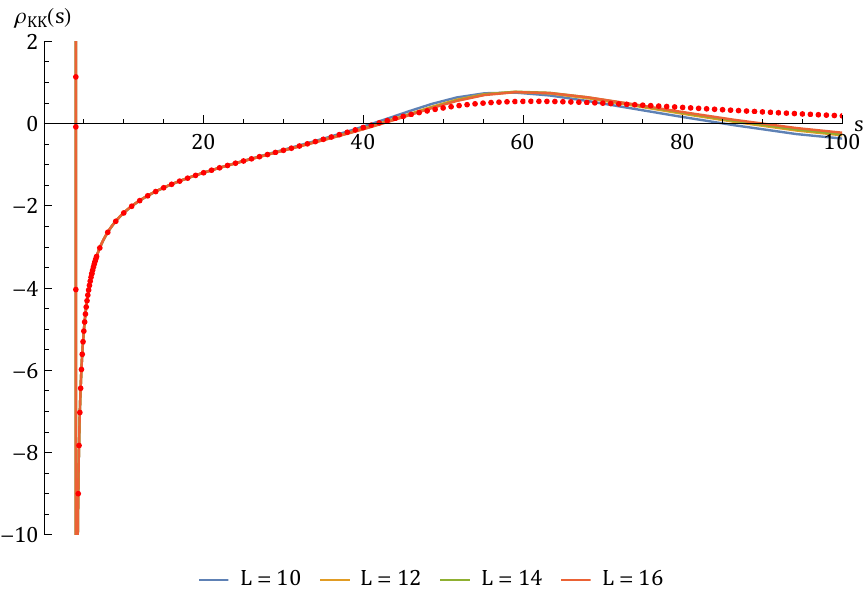}
      \caption{} \label{rhoa0a2fixedlambdamincomparisonLmax10Nmax10}
     \end{subfigure}
     \begin{subfigure}{0.49\textwidth}
         \centering
         \includegraphics[width=1\textwidth]{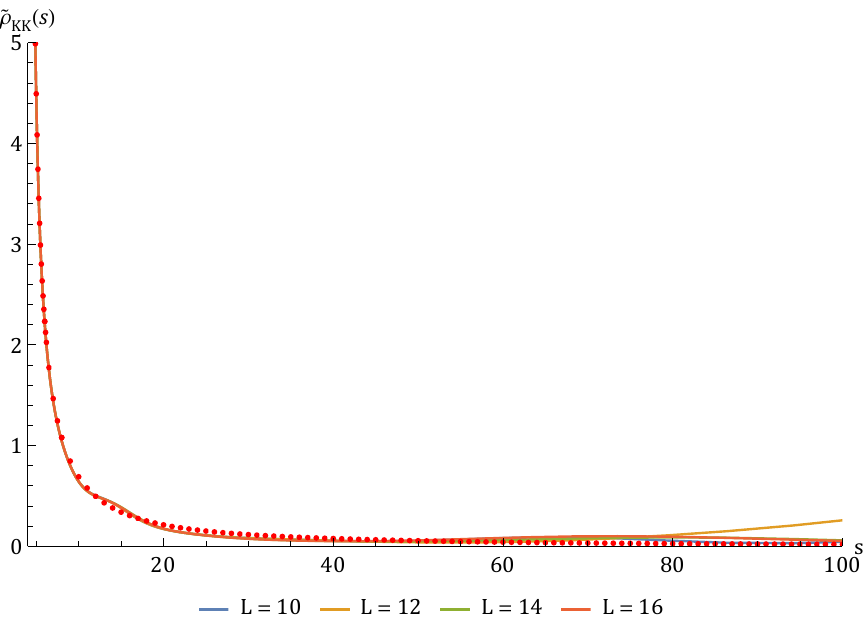}
     \caption{}  \label{Inverserhocomparisonalpha0max}
     \end{subfigure} 
      \caption{Left: Comparison plot of $\rho_{KK}(s)$ for $\lambda_{min}$ with $a_0$ and $a_2$ fixed to the pion values. Right: Comparison plot for $\tilde{\rho}_{KK}(s)=\frac{2}{\rho_{KK}(s)\sqrt{s-4}}$ for $\lambda_{max}$.}
\end{figure}

In order to compare the two methods, we minimize the quartic coupling $\lambda$ while setting $a_0$ and $a_2$ to pion values. We see good agreement between the two approaches for a wide range of parameters (see table \ref{Tabcomp}). It is also interesting to note that $D_B$ has better convergence for the position of the second zero of $\rho_{KK}(s)$, $s_c^{(2)}$, in spite of the significantly less number of constraints imposed (table \ref{Tabcomp}). While for $D_B$-bootstrap, we use $s_{max} \approx 4950$, for $\rho_W$-bootstrap, we impose non-linear unitarity upto $s_{max} \approx 5260$ and the improved positivity constraints up to $s_{max}\approx 10^9$. The $\rho_{KK}(s)$ for $\lambda_{min}$ with $a_0$ and $a_2$ fixed to the pion values and $\tilde{\rho}_{KK}(s)=\frac{2}{\rho_{KK}(s)\sqrt{s-4}}$ for $\lambda_{max}$ have been compared in fig(\ref{rhoa0a2fixedlambdamincomparisonLmax10Nmax10}) and fig(\ref{Inverserhocomparisonalpha0max}) with roughly the same number of parameters and we observe close agreement in their behaviors. 

\begin{figure}[H]
     \centering
     \begin{subfigure}{0.4882\textwidth}
         \centering
         \includegraphics[width=1\textwidth]{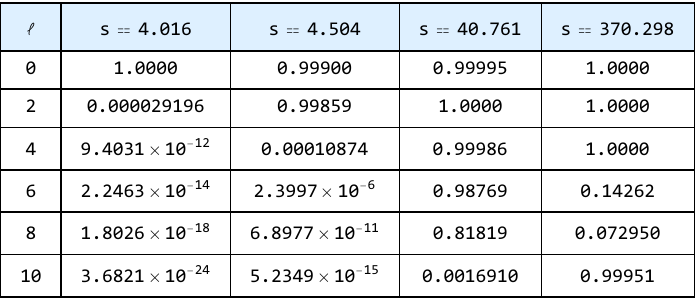}
      \caption{} \label{inelasticity}
     \end{subfigure}
     \begin{subfigure}{0.502\textwidth}
         \centering
         \includegraphics[width=1\textwidth]{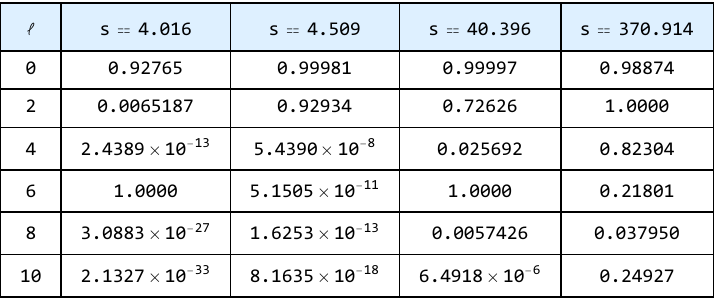}
     \caption{}  \label{inelasticityMiro}
     \end{subfigure} 
      \caption{The quantity $\frac{|f_\ell(s)|^2}{\text{Im}f_\ell(s)}$ is equal to 1 for elastic unitarity and less than 1 otherwise. Left: Values of $\frac{|f_\ell(s)|^2}{\text{Im}f_\ell(s)}$ for the amplitude with minimum of the quartic coupling with $a_0$ and $a_2$ fixed to the pion values using $D_{B}$-bootstrap with unitarity imposed up to $L=10$ and $s_{max}\approx4950$. Right: Same values using the methods of \cite{Miro} for $L_{max}=10,\,N_{max}=10$ and $s_{max} \approx 5260$.}
      \label{inelasticityComp}
\end{figure}

Finally, we also compare the amount of inelasticity in the amplitudes using both approaches as summarized in table \ref{inelasticityComp}. While the spin-0 results are comparable, there are interesting differences for higher spins, which do not appear to affect the final outcome. It will be interesting to incorporate elastic unitarity between $4\leq s \leq 16$ to see how the results change.

\section{$\rho_{KK}$ along the leaf}
\label{LeafRhos}
\begin{figure}[H]
     \centering
     \includegraphics[width=0.55\textwidth]{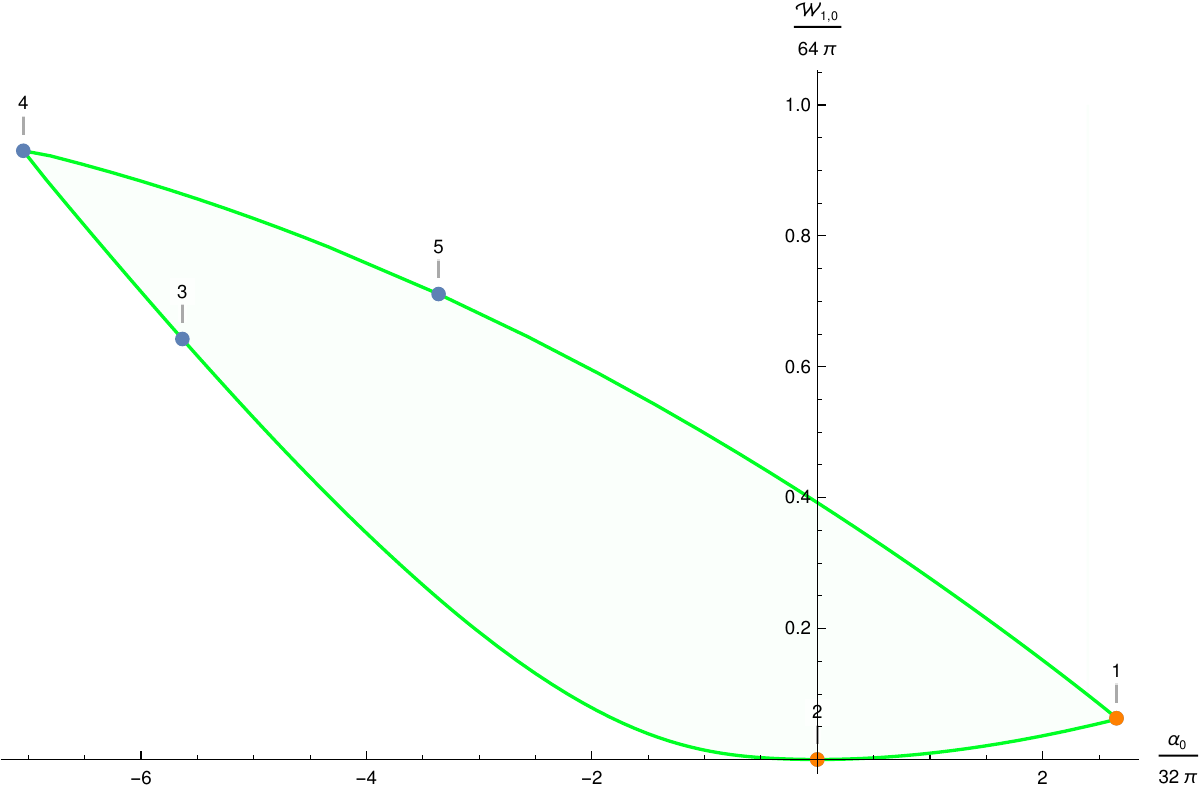}
     \caption{Points on the leaf plot for which we plot $\rho_{KK}(s)$ in figure \ref{RhoLeaf}. Orange points don't show a change in sign beyond (near) the threshold. Blue points do. }
     \label{Leaf}
\end{figure}

\begin{figure}[H]
     \centering
     \includegraphics[width=0.7\textwidth]{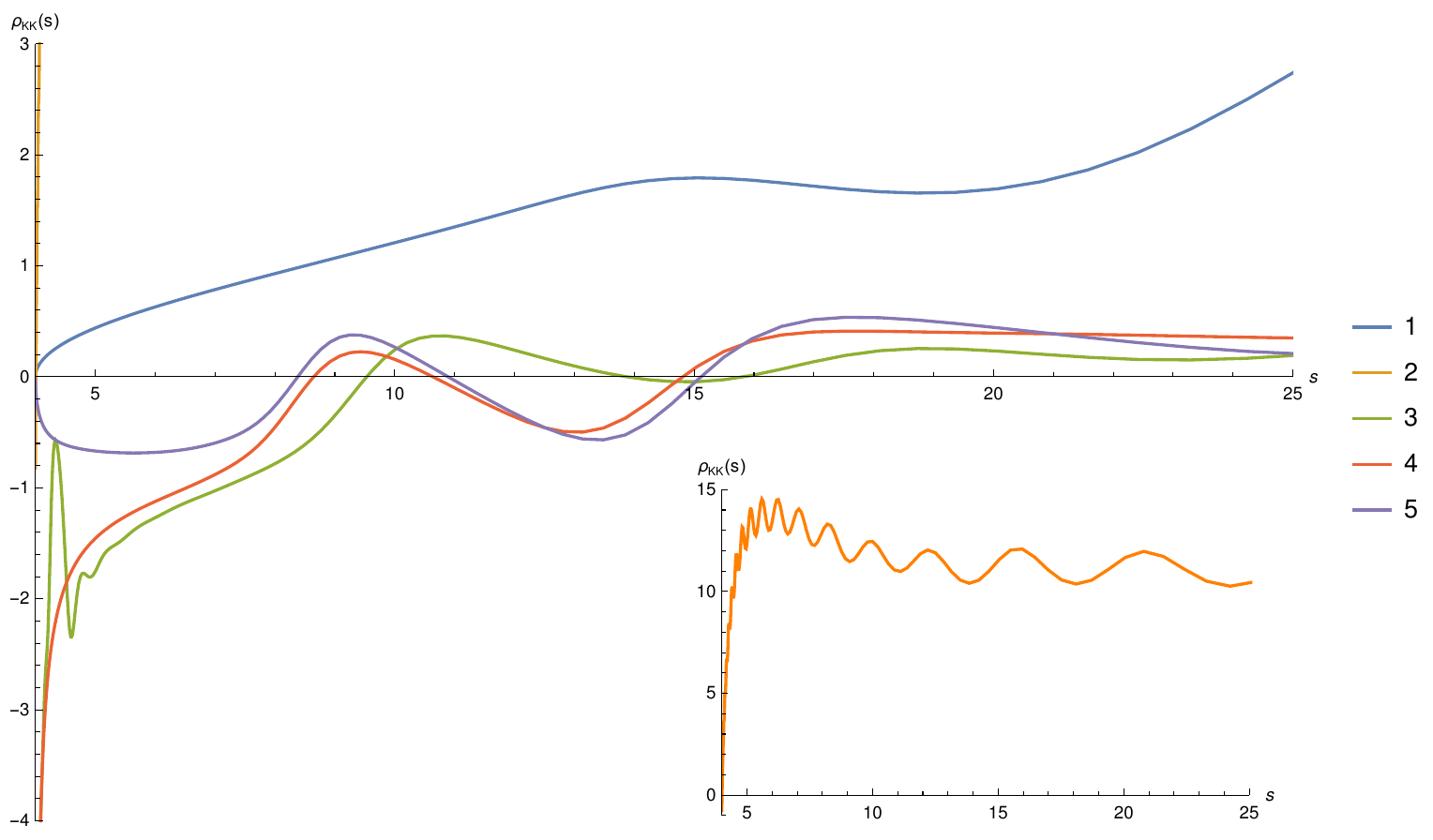}
     \caption{$\rho_{KK}(s)$ for points on the leaf. The mini plot shows $\rho_{KK}(s)$ for point 2 separately as it shoots up very fast and is barely visible in the main plot.}
     \label{RhoLeaf}
\end{figure}

In this section, we plot $\rho_{KK}$ for some of the points along the leaf plot (see figure 2 of \cite{Miro})\footnote{We thank Joan Miro and Mehmet Gumus for sharing their data.}. Figure \ref{Leaf} shows the points for whch we plot $\rho_{KK}(s)$ in figure \ref{RhoLeaf}. 

 We notice that the amplitude corresponding to the maximum quartic coupling (curve 1) shows no change in sign anywhere. The amplitude corresponding to the origin of the leaf (curve 2) shows interesting features. As can be (barely) seen in the main plot, it shoots up very fast near $s =4$. So we plot it in a mini-plot inside the main plot and see that after shooting up, it starts oscillating sideways. The amplitude corresponding to the minimum quartic coupling (curve 4) and its adjacent points (curves 3 and 5) show somewhat similar behaviours. Interestingly, they show three changes in sign and two peaks before asymptoting, compared to the amplitude with minimum $\lambda$ and $a_2 = a_2^{(pion)}$ which shows only one peak and two zeroes - $s_c^{(1)}$ and $s_c^{(2)}$.

 \section{Computation of phase shifts}
\begin{figure}[H]
     \centering
     \includegraphics[width=1\textwidth]{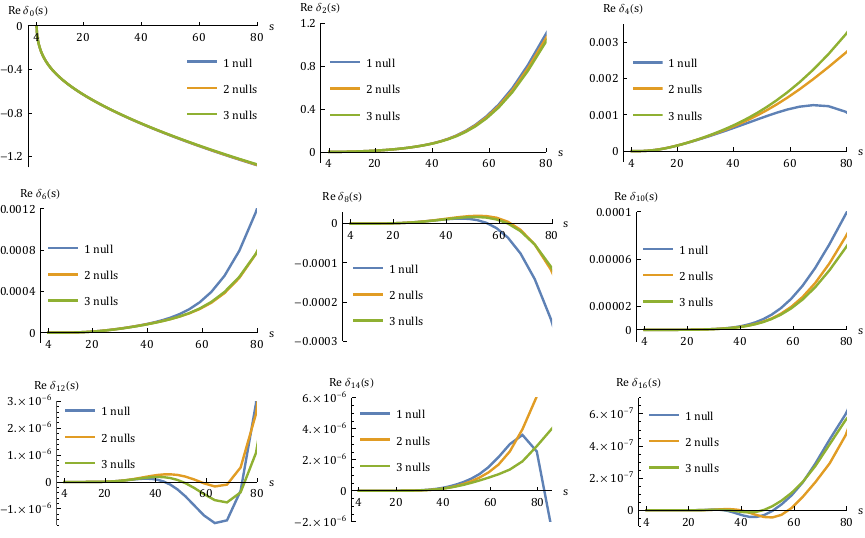}
     \caption{Behaviour of $\text{Re}~\delta_{\ell}(s)$ as the number of null constraints is increased for the amplitude with minimum $\lambda$ and $a_0$, $a_2$ fixed to pion values}
     \label{phaseshifts}
\end{figure}

\begin{figure}[H]
     \centering
     \begin{subfigure}{0.49\textwidth}
         \centering
         \includegraphics[width=0.9\textwidth]{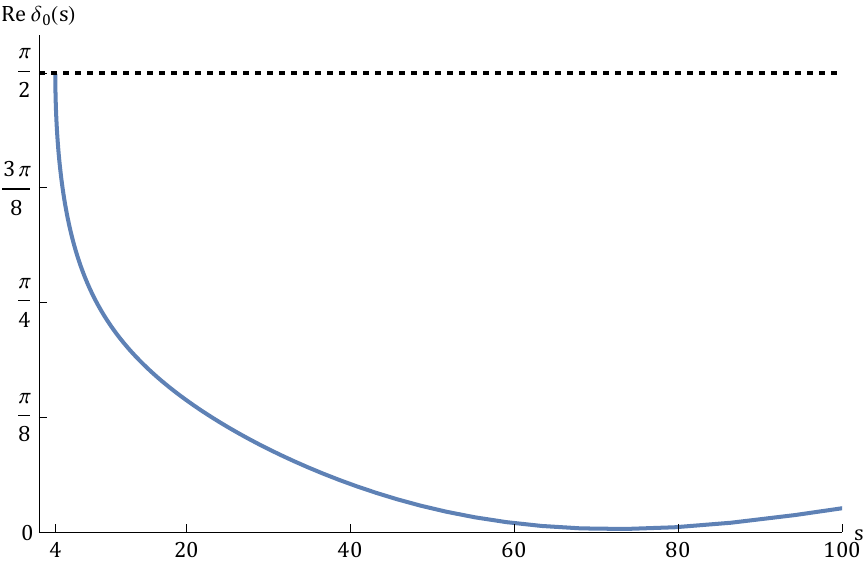}
      \caption{$\text{Re}~\delta_{0}(s)$ for $\lambda_{max}$.}
     \end{subfigure}
     \begin{subfigure}{0.49\textwidth}
         \centering
         \includegraphics[width=0.95\textwidth]{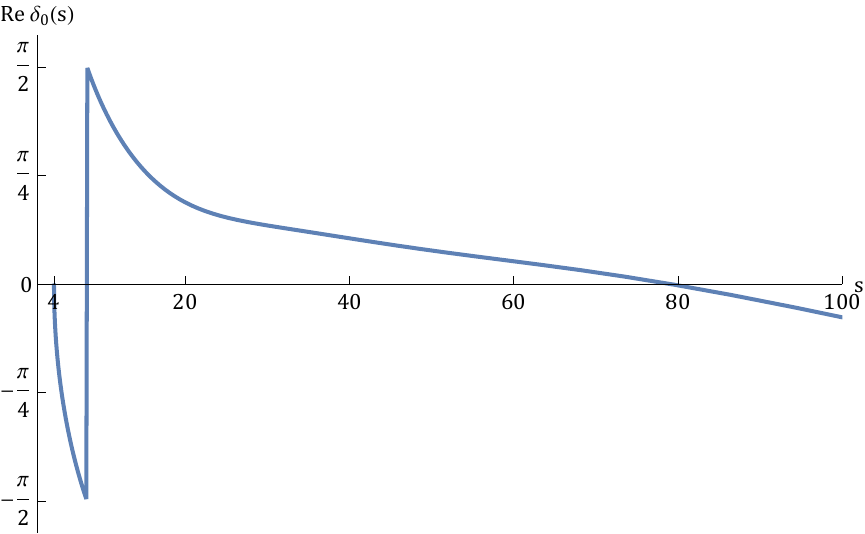}
     \caption{$\text{Re}~\delta_{0}(s)$ for $\lambda_{min}$.}  
     \end{subfigure} 
      \caption{Behaviour of $\text{Re}~\delta_{0}(s)$ for the amplitudes with minimum and maximum $\lambda$ with 1 null constraint.}
      \label{spin0phaseshiftLambdamaxmin}
\end{figure}

 Defining $S_{\ell}(s) = 1 + 2i f_{\ell}(s)$ allows us to express the unitarity condition $|f_{\ell}(s)|^2 \le \text{Im}f_{\ell}(s) \le 1$ simply as $|S_{\ell}(s)| \le 1$. We can the define the phase shifts $\delta_{l}(s)$ as $S_{\ell}(s) = e^{2i \delta_{l}(s)}$. $\delta_{l}(s)$ are purely real only in the elastic region  $4 \le s \le 16$; in general, they are complex with a non-negative imaginary part to satisfy unitarity. Since we do not impose elastic unitarity, for the amplitudes we construct, the phase shifts can always be complex. \\
 In the figure \eqref{phaseshifts}, for the amplitude with minimum $\lambda$, and $a_0$ and $a_2$ fixed to pion values, we show how $\text{Re}~\delta_{\ell}(s)$ behaves as we increase the total number of null constraints imposed. \\
 We notice that $\text{Re}~\delta_{\ell}(s)$ converges well with just 3 null constraints ($\mathcal{W}_{1,m},~ m = 2,3,4$) upto spin 10. Beyond that, we need more null constraints. We have checked convergence upto $\text{Re}~\delta_{16}(s)$ has requires 5 null constraints ($\mathcal{W}_{1,m},~ m = 2,3,4,5,6$).\\
In the figure \eqref{spin0phaseshiftLambdamaxmin}, we also show how $\text{Re}~\delta_{0}(s)$ behaves for the amplitudes with maximum and minimum $\lambda$ respectively. 

\end{document}